\documentclass{fundam}
\pdfoutput=1

 \publyear{24}
 \papernumber{2197}
 \volume{192}
 \issue{3-4}

   \finalVersionForARXIV

\usepackage{amssymb,amsmath}
\usepackage{bbm}
\usepackage{stmaryrd}
\usepackage{varwidth}
\usepackage{siunitx}
\usepackage{booktabs}

\usepackage{tabularx}
\usepackage{arydshln}
\usepackage{marvosym}
\usepackage{xspace}

\newcommand{\Tina}{\texttt{Tina}\xspace}
\newcommand{\prule}[1]{\textsc{({#1})}\xspace}

\renewcommand{\geq}{\geqslant}

\newcommand{\pre}{\mathrm{Pre}\xspace}
\newcommand{\post}{\mathrm{Post}\xspace}

\newcommand{\fv}[1]{\ensuremath{\mathrm{FV}({#1})}}

\newcommand{\maseq}[1]{\underline{#1}}

\newcommand{\cec}{\mathop{\langle C_1 E C_2 \rangle}}
\newcommand{\tleft}[1]{\acute{#1}}
\newcommand{\tadd}{T^{+}}
\newcommand{\tsub}{T^{-}}

\let\s\sigma

\newcommand{\sprod}{\mathop{\|}}

\newcommand{\comma}{\wedge}

\newcommand{\Nat}{\mathbb{N}}

\let\imply\implies
\newcommand{\st}{\ .\ }

\newcommand{\subst}[2]{\{#1 \leftarrow #2\}}

\newcommand{\ftrans}[1]{\wtrans{#1\rangle}}

\newcommand{\eequiv}[1][E]{\mathop{\equiv_{#1}\,}}

\newcommand{\reduc}{\equiv}
\newcommand{\eabs}{\sqsubseteq}

\newcommand{\preduc}{\approxeq}
\newcommand{\pabs}{\preceq}

\newcommand*{\defeq}{\triangleq}
\let\eqdef\defeq
\makeatletter
\def \rightarrowfill{\m@th\mathord{\smash-}\mkern-6mu%
  \cleaders\hbox{$\mkern-2mu\mathord{\smash-}\mkern-2mu$}\hfill
  \mkern-6mu\mathord\rightarrow}
\makeatother
\makeatletter
\def \Rightarrowfill{\m@th\mathord{\smash-}\mkern-6mu%
  \cleaders\hbox{$\mkern-2mu\mathord{\smash-}\mkern-2mu$}\hfill
  \mkern-6mu\mathord\Rightarrow}
\makeatother
\makeatletter
\def \rightarrowfill{\m@th\mathord{\smash-}\mkern-6mu%
  \cleaders\hbox{$\mkern-2mu\mathord{\smash-}\mkern-2mu$}\hfill
  \mkern-6mu\mathord\rightarrow}
\makeatother
\makeatletter
\def \Rightarrowfill{\m@th\mathord{\smash=}\mkern-6mu%
  \cleaders\hbox{$\mkern-2mu\mathord{\smash=}\mkern-2mu$}\hfill
  \mkern-6mu\mathord\Rightarrow}
\makeatother
\makeatletter
\def \midrightarrowfill{\m@th\mathord{\smash{\raisebox{.2ex}{$\scriptscriptstyle\mid$}}\!\!\,-}\mkern-6mu%
  \cleaders\hbox{$\mkern-2mu\mathord{\smash-}\mkern-2mu$}\hfill
  \mkern-6mu\mathord\rightarrow}
\makeatother
\makeatletter
\def \midRightarrowfill{\m@th\mathord{\smash{\raisebox{.1ex}{$\scriptstyle\mid$}}\!\!\!=}\mkern-6mu%
  \cleaders\hbox{$\mkern-2mu\mathord{\smash=}\mkern-2mu$}\hfill
  \mkern-6mu\mathord\Rightarrow}
\makeatother
\newcommand{\overstackrel}[2]{\mathrel{\mathop{#1}\limits^{#2}}}
\newcommand{\trans}[1]{\mathbin{\smash[t]{\overstackrel{\rightarrowfill}{\ #1\ }}}}
\newcommand{\wtrans}[1]{\mathbin{\smash[t]{\overstackrel{\Rightarrowfill}{\ #1\ }}}}

\renewcommand{\vec}[1]{\boldsymbol{#1}}
\newcommand{\sem}[1]{\llbracket {#1} \rrbracket}

\usepackage{tikz}
\usetikzlibrary{arrows}
\usetikzlibrary{decorations.markings}
\usetikzlibrary{shadows}
\usepackage{subcaption}
\tikzset{
  big stealth/.style={
    decoration={markings,mark=at position -(0.1pt) with {\arrow[scale=2*\scale]{stealth}}},
    postaction={decorate},
    shorten >=0.4pt}}
\tikzset{
  big ring/.style={
    decoration={markings,mark=at position -(0.1pt) with {\arrow[scale=1.5*\scale]{o}}},
    postaction={decorate},
    shorten >=8pt*\scale}}
\tikzset{
  big disc/.style={
    decoration={markings,mark=at position -(0.1pt) with {\arrow[scale=1.5*\scale]{*}}},
    postaction={decorate},
    shorten >=8pt*\scale}}
\tikzset{
  big box/.style={
    decoration={markings,mark=at position -(0.1pt) with {\arrow[scale=1.5*\scale]{open square}}},
    postaction={decorate},
    shorten >=8pt*\scale}}
\tikzset{
  big tile/.style={
    decoration={markings,mark=at position -(0.1pt) with {\arrow[scale=1.5*\scale]{square}}},
    postaction={decorate},
    shorten >=8pt*\scale}}

\usetikzlibrary{shapes,backgrounds,decorations.text,patterns}
\usetikzlibrary{decorations.pathreplacing}

\tikzstyle{state}=[circle, very thick, fill, top color=white, bottom color=white, draw=black, minimum size=40pt]
\tikzstyle{place}=[circle, very thick, fill, top color=white, bottom color=white, draw=black, minimum size=40pt]
\tikzstyle{bplace}=[circle, very thick, fill, top color=cyan, fill
opacity=0.2, text opacity=1, bottom color=white, draw=cyan, minimum size=40pt]
\tikzstyle{vplace}=[circle, very thick, fill, top color=violet, fill
opacity=0.2, text opacity=1, bottom color=white, draw=violet, minimum size=40pt]
\tikzstyle{trans}=[rectangle, very thick, fill, top color=white, bottom color=white, draw=black, minimum size=32pt]
\tikzstyle{btrans}=[rectangle, very thick, fill, top color=cyan, fill
opacity=0.2, text opacity=1, bottom color=white, draw=cyan, minimum size=32pt]
\tikzstyle{vtrans}=[rectangle, very thick, fill, top color=violet, fill
opacity=0.2, text opacity=1, bottom color=white, draw=violet, minimum size=32pt]
\tikzstyle{bvtrans}=[rectangle, fill, bottom color=violet, top color=cyan, fill
opacity=0.2, text opacity=1, minimum size=32pt]
\tikzstyle{arc}=[thick, big stealth, black]
\tikzstyle{barc}=[thick, big stealth, cyan]
\tikzstyle{varc}=[thick, big stealth, violet]
\tikzstyle{read}=[thick, big disc, black]
\tikzstyle{inhibitor}=[thick, big ring, black]
\tikzstyle{stopwatch}=[thick, big tile, black]
\tikzstyle{stopwatchinhibitor}=[thick, big box, black]
\tikzstyle{priority}=[thick, big stealth, orange]
\tikzstyle{enabling}=[thick, big disc, orange]
\tikzstyle{disabling}=[thick, big ring, orange]
\tikzstyle{token}=[circle, fill, draw=black, minimum size=4pt]
\tikzstyle{btoken}=[circle, fill, top color=cyan, fill opacity=0.5, text opacity=1, bottom color=cyan, draw=cyan, minimum size=4pt]
\tikzstyle{glob-options}=[label
distance=6pt*\scalenodes*\scale,x=1pt,y=-1pt,scale=\scale,every
node/.style={transform shape}, baseline=(current bounding box.center)]
\tikzstyle{virtual}=[circle, draw=white, minimum size=20pt]

\def\scale{0.57}
\def\scalenodes{1.0}

\newcommand*{\strans}[1]{\mathbin{\smash[t]{\overstackrel{\rightarrowfill}{\ #1\ }}}}
\newcommand*{\inv}{\mathrm{Inv}}

\newenvironment{recalldef}[2]{
  \begin{quote}\textbf{Definition~\ref{#1} \emph{(#2)}}\\
}
{\end{quote}}

\newenvironment{recallthm}[2]{
  \begin{quote}\textbf{Theorem~{#1} \emph{(#2)}}\\
}
{\end{quote}}

\usepackage[ruled,lined]{algorithm2e}
\usepackage{graphicx}
\usepackage{tikz}
\usetikzlibrary{automata, positioning, arrows}

\usepackage[T1]{fontenc}
\usepackage{xcolor}
\usepackage{hyperref}
\hypersetup{
  colorlinks=true
  citecolor=violet,
  linkcolor=red,
  urlcolor=blue
  }

\usepackage{url} 

\begin{document}

\title{On the Complexity of Proving Polyhedral Reductions}

\author{Nicolas Amat\thanks{Address for correspondence:  IMDEA Software Institute
Campus Montegancedo s/n, 28223 Pozuelo de Alarcon, Madrid, Spain}
\\
IMDEA Software Institute\\
Madrid, Spain\\
nicolas.amat{@}imdea.org
\and Silvano Dal Zilio,\ Didier Le Botlan\\
LAAS-CNRS\\
Universit\'e de Toulouse, INSA, CNRS\\
Toulouse, France}

\maketitle

\runninghead{N. Amat et al.}{On the Complexity of Proving Polyhedral Reductions} 

\begin{abstract}
  We propose an automated procedure to prove polyhedral abstractions (also
  known as polyhedral reductions)
  for Petri
  nets. Polyhedral abstraction is a new type of state space equivalence, between
  Petri nets, based on the use of linear integer constraints between the marking of places.
  In
  addition to defining an automated proof method, this paper aims to better characterize
  polyhedral reductions,
  and to give an overview of their application to reachability
  problems.

  Our approach relies on encoding the equivalence problem into a set of SMT formulas whose
  satisfaction implies that the equivalence holds. 
  The difficulty, in this
  context, arises from the fact that we need to handle infinite-state systems.
  For completeness, we exploit a connection with a class of Petri nets, called
  flat nets, that have Presburger-definable reachability sets.
  We have
  implemented our procedure, 
  and we illustrate its use on several examples.
\end{abstract}

\begin{keywords}
  Automated reasoning, Abstraction techniques, Reachability problems, Petri nets
\end{keywords}

\section{Introduction}

Our work is related with a new abstraction
technique for Petri nets~\cite{berthomieu2018petri,berthomieu_counting_2019} based on
a combination of structural
reductions~\cite{berthelot_checking_1985,berthelot_transformations_1987} with the use of
 linear constraints between
the marking of places. The
idea is to compute reductions of the form $(N, E, N')$, where: $N$ is an initial
net (that we want to analyze); $N'$ is a residual net (hopefully much simpler
than $N$); and $E$ is a Presburger predicate. The idea is to preserve enough
information in $E$ to rebuild the reachable markings of $N$, knowing only the
ones of $N'$.
We refined this concept into a new abstraction, called \emph{polyhedral
abstraction}~\cite{fi2022,sttt2022}, in reference to ``polyhedral models'' used in program optimization
and static analysis~\cite{besson_polyhedral_1999,feautrier_automatic_1996}.
Indeed, like in these works, we propose an algebraic representation of
state spaces using solutions to
linear constraints. We use the term \emph{abstraction} to refer to the fact
that we exploit an equivalence relation that provides a mapping between an initial and a target model.
We shall also often use the name \emph{polyhedral reduction} for the same concept, interchangeably,
due to the fact that many of the abstraction rules we use in practice derives from (Petri net)
structural reductions.

We implemented our
approach into two independent symbolic model-checkers developed by our team: \texttt{Tedd}, a
tool based on Hierarchical Set Decision Diagrams
(SDD)~\cite{thierry2009hierarchical}, part of the \Tina
toolbox~\cite{tinaToolbox}; and \texttt{SMPT}~\cite{smpt,fm2023}, an SMT-based
model-checker focused on reachability problems~\cite{tacas2022}. Both tools
demonstrated the effectiveness of polyhedral reductions by
achieving good rankings in both the StateSpace and Reachability examinations of the
Model Checking Contest~\cite{mcc2019}, an international competition for model-checking tools.

Our approach has several positive features. In particular, it does not impose
restrictions on the syntax of nets, such as constraints on the weights of arcs,
and it can be transparently applied to unbounded nets. In practice, we can often
reduce a Petri net $N$ with $n$ places (from a {high dimensional} space) into a
residual net $N'$ with far fewer places, say $n'$ (in a lower-dimensional
space). More formally, with our approach, we can represent the state space of
$N$ as the ``inverse image'', by the Presburger predicate $E$, of the state
space of $N'$ (a subset of vectors in dimension $n'$), which can result in a
very compact representation of the reachability set. This problem shares some
similarities with the question of whether we can precisely characterize the
reachability set of a net using a formula in Presburger arithmetic. A connection
we will further develop. An important distinction is that we use ``Presburger
relations'' to relate the reachability set of two nets, as an equivalence,
rather than to abstract a single state space. One of the goals of our work is to
give decidability results about this equivalence, and to find ways to
automatically check when an equivalence judgment is true.

We define this notion of equivalence using a new relation, $N \reduc_{E}
N'$, called \emph{polyhedral abstraction equivalence} (or just \emph{polyhedral
equivalence} for short). We should also often use the term \emph{$E$-abstraction
equivalence} to emphasize the importance of the linear predicate $E$. This
equivalence plays a central role in many of our results, as well as it provides the basis to
formally define polyhedral reductions.

We prove that deciding the correctness of our original notion of
equivalence, see Sect.~\ref{sec:polyabs}, is undecidable (Theorem~\ref{th:undecidability-E-equivalence}). This
decidability result is not surprising since most equivalence problems on Petri
nets are undecidable~\cite{esparza1994decidability,esparza1998decidability}.
Indeed, polyhedral equivalence is by essence related to the \emph{marking
equivalence} problem, which amounts to deciding if two Petri nets with the same
set of places have the same reachable markings; a problem proved undecidable by
Hack~\cite{hack1976decidability}. Also, polyhedral equivalence (such as marking
equivalence) entails trace equivalence, another well-known undecidable
equivalence problem when we consider general Petri
nets~\cite{hirshfeld1994petri,hack1976decidability}.

Although this may appear contradictory, we prove that the equivalence problem
becomes decidable when we consider a slightly different, and in some sense more
general, equivalence relation between \emph{parametric} Petri nets.
In this context, we use the term \emph{parametric} to stress the fact that we manipulate
semilinear sets of markings, meaning sets that can be defined using a Presburger
arithmetic formula $C$. In particular, we reason about parametric nets $(N, C)$,
instead of marked nets $(N, m_0)$, with the intended meaning that all markings
satisfying $C$ are potential initial markings of $N$. We also define an extended
notion of polyhedral equivalence between parametric nets, denoted $(N_1, C_1)
\preduc_E (N_2, C_2)$, whereas our original definition~\cite{pn2021,fi2022} was
between marked nets only (see Definition~\ref{def:eabs}).

We show that given a valid equivalence statement $(N_1, C_1) \preduc_E (N_2,
C_2)$, it is possible to derive a Presburger formula, in a constructive way,
whose satisfaction implies that the equivalence holds. We implemented this
procedure on top of an SMT-solver for Linear Integer Arithmetic (LIA) and show
that our approach is applicable in practice (Sect.~\ref{sec:examples}). Even if
we prove that this problem is decidable (see
Theorem~\ref{th:parametric-E-abstraction-decidable}), our implementation is only
a semi-decision procedure since we rely on the external tool \texttt{FAST},
which may not terminate if the equivalence does not hold. If anything, it makes
the fact that we may translate our problem into Presburger arithmetic quite
remarkable.

\subsubsection*{Description of our approach}

Our approach can be summarized as follows. We start from
an initial net $(N_1, C_1)$ and derive a polyhedral equivalence
$(N_1, C_1) \preduc_E (N_2, C_2)$ by applying a set of
\emph{abstraction laws} in an iterative and compositional way. Finally, we
solve a reachability problem about $N_1$ by transforming it into a
reachability problem about net $N_2$, which should hopefully be easier to
check. A large number of the laws we implement in our tools derive from structural
reduction rules~\cite{berthelot_transformations_1987}, or are based on
the elimination of redundant places and transitions, with the goal to
obtain a ``reduced'' net $N_2$ that is smaller than $N_1$.

We also implement several other kinds of abstraction rules---often
subtler to use and harder to prove correct---which explains why we want
machine checkable proofs of equivalence. For instance, some of our rules
are based on the identification of Petri nets subclasses in which the set
of reachable markings equals the set of potentially reachable ones, a
property we call the PR-R equality
in~\cite{DBLP:journals/corr/abs-2006-05600,DBLP:journals/corr/abs-2005-04818}. We
use this kind of rules in the example of the ``SwimmingPool'' model of
Fig.~\ref{fig:Swimming_pool}, a classical example of Petri net often
used in case studies (see e.g.~\cite{10.1007/3-540-48320-9_14}).

\begin{figure}[ht]
\vspace*{-2mm}
  \centering \includegraphics[width=0.70\linewidth]{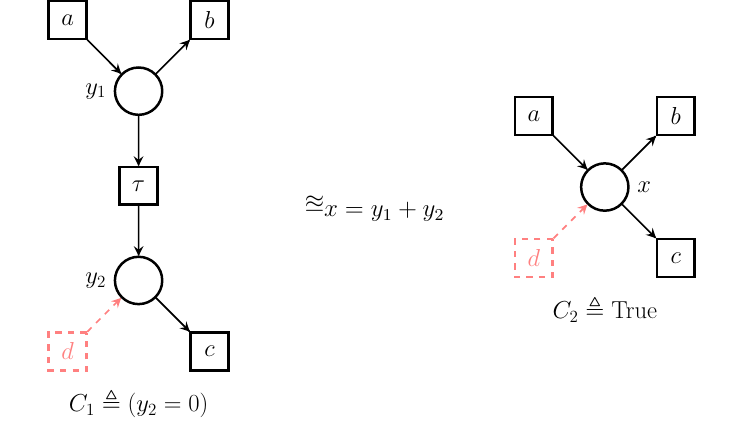}
  \caption{Equivalence rule \prule{concat},
    $(N_1, C_1) \preduc_E (N_2, C_2)$, between nets $N_1$ (left) and
    $N_2$ (right), for the relation
    $E \defeq (x = y_1 + y_2)$.\label{fig:concat}}
\end{figure}
\begin{figure}[!h]
\vspace*{-2mm}
  \centering
  \includegraphics[width=0.78\textwidth]{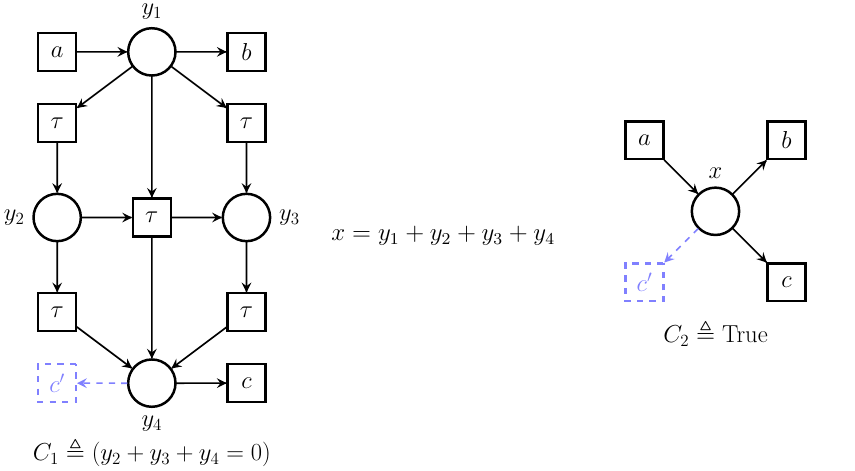}
  \caption{Equivalence rule \prule{magic}.\label{fig:magic}}\vspace*{-4mm}
\end{figure}

\medskip
We give an example of a basic {abstraction law} in
Fig.~\ref{fig:concat}, with an instance of rule \prule{concat} that
allows us to fuse two places connected by a direct, silent transition.
We give another example with \prule{magic}, in Fig.~\ref{fig:magic},
which illustrates a more complex agglomeration rule, and refer to
other examples in Sect.~\ref{sec:examples}.

The parametric net $(N_1, C_1)$ (left of Fig.~\ref{fig:concat}) has a
condition which entails that place $y_2$ should be empty initially
($y_2 = 0$), whereas net $(N_2, C_2)$ has a trivial constraint, which
can be interpreted as simply $x \geq 0$. We can show (see
Sect.~\ref{sec:strong_equivalence}) that nets $N_1$ and $N_2$ are
$E$-equivalent, which amounts to prove that any marking
$(y_1: k_1,\ y_2: k_2)$ of $N_1$, reachable by firing a transition
sequence $\sigma$, can be associated with the marking
$(x : k_1 + k_2)$ of $N_2$, also reachable by the same firing
sequence. Actually, we prove that this equivalence is sound when no
transition can input a token directly into place $y_2$ of $N_1$. This
means that the rule is correct in the absence of the ``dashed'' transition
(with label $d$), but that our procedure should flag the rule as
unsound when transition $d$ is present.

\medskip
The results presented in this paper provide an automated technique for
proving the correctness of polyhedral abstraction laws. This helps us
gain more confidence on the correctness of our tools and is also
useful if we want to add new abstraction rules. Indeed, up until now,
all our rules where proven using ``manual theorem proving'', which can
be tedious and error-prone.

Incidentally, the theory we developed for this paper also helped us
gain a better understanding of the constraints necessary when
designing new abstraction laws. A critical part of our approach relies
on the ability, given a Presburger predicate $C$, to encode the set of
markings reachable from $C$ by firing only silent transitions,
that we denote $\tau^\star_C$ in the following. Our approach
draws a connection with previous
works~\cite{bardin2003fast,bardin_fast_2008,presburger_vass} that
study the class of Petri nets that have Presburger-definable reachability
sets; also called \emph{flat nets}. We should also make use of a tool
implemented by the same authors, called \texttt{FAST}, which provides
a method for representing the reachable set of flat nets.
Basically, we gain the insight that polyhedral reductions provide a
way to abstract away (or collapse) the sub-parts of a net that are
flat. Note that our approach may work even though the
reachability set of the whole net is not semilinear, since only the
part that is abstracted must be flat. We also prove that when
$(N_1,C_1) \preduc_E (N_2,C_2)$ then necessarily the sets
$\tau^\star_{C_1}$ and $\tau^\star_{C_2}$ are semilinear.

\subsubsection*{Outline and contributions}
The paper is organized as follows. We define our central notion of
\emph{parametric polyhedral abstraction} in Sect.~\ref{sec:strong_equivalence}
and prove several of its properties in Sect.~\ref{sec:construction-laws}. In
particular, we prove that polyhedral abstraction is a congruence, and that it is
preserved when ``duplicating labeled transitions''. These properties mean that
every abstraction law we prove can be safely applied in every context, and that
each law can be used as a ``rule schema''. Our definition relies on a former
notion of polyhedral equivalence, that we recall in
Sect.~\ref{sec:petri_polyhedral}, together with a quick overview of our
notations. We describe our proof procedure in Sect.~\ref{sec:procedure}, which
is defined as the construction of a set of four \emph{core requirements}, each
expressed as separate quantified LIA formulas. A key ingredient in this
translation is to build a predicate, $\tau^\star_C$, which encodes the markings
reachable by firing only the silent transitions of a net. We defer the
definition of this predicate until Sect.~\ref{sec:silent_transition_relation},
where we show how it can be obtained using the output of the \texttt{FAST} tool.
From this procedure, we prove that our problem is decidable in
Sect.~\ref{sec:decidability}, and we extend our automated procedure in
Sect.~\ref{sec:checking-state-space-partition} to the check of state space
partition that is prerequisite for model counting. We conclude by presenting the
results obtained with a new tool implementing our approach, called
\texttt{Reductron}, on some concrete examples. First, in
Sect.~\ref{sec:debugging} by showing how our tool can be used to ``debug''
incorrect reduction rules, and in Sect.~\ref{sec:examples} by providing
quantitative information on its performance for our set of reduction rules.

Many results and definitions have already been presented in a shorter version of
the paper~\cite{amat_automated_2023}. This extended version contains several
additions. First, we have added the full proofs of all the results given in our
work and added a new fundamental result, namely the decidability of checking
parametric equivalence (Sect.~\ref{sec:decidability}). We also give a more
precise proof for the undecidability of checking ``regular'' polyhedral
equivalence. This paper also contains new theoretical results, such as an
automatic method for checking when the equivalence defines a partition of the
state space (Sect.~\ref{sec:checking-state-space-partition}). Finally, we added a new section with experimental results
about the performance of our tool (Sect.~\ref{sec:examples}) and its use for debugging
problematic reduction rules (Sect.~\ref{sec:debugging}).


\section{Petri nets and polyhedral abstraction}
\label{sec:petri_polyhedral}

In this section, we briefly introduce Petri nets and our concept of polyhedral
reduction. While we tried to make the presentation as self-contained as
possible, we assume that the reader has some familiarity with basic Petri net
theory.

\subsection{Petri nets}

A Petri net is a tuple $(P, T, \pre, \post)$, where $P \defeq \{p_1, \dots, p_n\}$ is a
finite set of places, $T \defeq \{t_1, \dots, t_k\}$ is a finite set of transitions
(disjoint from $P$), and $\pre : T \rightarrow (P \rightarrow
\mathbb{N})$ and $\post : T \rightarrow (P \rightarrow \mathbb{N})$ are the pre-
and post-condition functions (also known as the flow functions of the net). A
state of a net, also called a marking, is a mapping $m : P \rightarrow
\mathbb{N}$ (also denoted $\Nat^P$) that assigns a number of tokens, $m(p)$, to each place $p$ in $P$. A
marked net $(N, m_0)$ is a pair consisting of a net, $N$, and an initial marking,
$m_0$.
In the following, we will often consider that each transition is labeled with a
symbol from an alphabet $\Sigma$. In this case, we assume that a net is
associated with a labeling function $l : T \to \Sigma \cup \{\tau\}$, where $\tau$
is a special symbol for the silent action. Every net has a default labeling
function, $l_N$, such that $\Sigma = T$ and $l_N(t) \defeq t$ for every transition $t
\in T$.

A transition $t \in T$ is enabled at a marking $m \in \Nat^P$ if $m(p) \geq
\pre(t, p)$ for all places $p \in P$, which we also write $m \geq
\pre(t)$, where $\geq$ represents component-wise comparison of the
markings. A marking $m' \in \Nat^P$ is reachable from a marking $m \in \Nat^P$
by firing transition $t$, denoted $(N,m) \trans{t} (N,m')$ or simply $m \trans{t} m'$ when $N$ is obvious from the context, if: (1) transition $t$ is
enabled at $m$, and (2) $m' = m - \pre(t) + \post(t)$. A firing
sequence $\varrho \defeq t_1, \dots, t_n \in T^*$ can be fired from $m$,
denoted $(N,m) \strans{\varrho} (N,m')$ or simply $m \strans{\varrho} m'$,
if there exist markings $m_0, \dots, m_n$ such that $m =
m_0$, $m' = m_n$, and $m_i \trans{t_{i+1}} m_{i+1}$ for all $i < n$.
We denote $R(N, m_0)$ the set of markings reachable from $m_0$ in $N$.

\subsubsection*{Observable sequences}

We can lift any labeling function $l : T \to \Sigma \cup \{\tau\}$
to a mapping of sequences from $T^*$ to $\Sigma^*$.
Specifically, we define inductively $l(\varrho.t) \defeq l(\varrho)$ if
$l(t) = \tau$ and $l(\varrho.t) \defeq l(\varrho).l(t)$ otherwise, where the dot operator
($.$) stands for concatenation, and $l(\epsilon) \defeq \epsilon$, where
$\epsilon$ is the empty sequence, verifying
$\epsilon.\sigma = \sigma.\epsilon = \sigma$ for any
$\sigma \in \Sigma^*$.
Given a sequence of labels $\sigma \in
\Sigma^*$, we write $(N,m) \wtrans{\sigma} (N, m')$ if there exists a firing
sequence $\varrho\in T^*$ such that $(N,m) \wtrans{\varrho} (N,m')$ and $\sigma
= l(\varrho)$. In this case, $\sigma$ is referred to as an \emph{observable sequence}
of the marked net $(N,m)$.
In some cases, we have to consider firing sequences that must not finish with $\tau$ transitions.
Hence, we define a relation $(N,m) \ftrans{\sigma} (N,m')$, written simply $m \ftrans{\s} m'$, as follows:
\begin{itemize}
\item $(N,m) \ftrans{\epsilon} (N,m)$ holds for all marking $m$.
\item $(N,m) \ftrans{\sigma.a} (N,m')$ holds for any markings $m,m'$ and $a,\sigma \in \Sigma \times \Sigma^*$,
  if there exists a marking $m''$ and a transition $t$ such that
  $l(t) = a$ and $(N,m) \wtrans{\sigma} (N,m'') \trans{t} (N,m')$.
\end{itemize}
It is immediate that $m \ftrans{\sigma} m'$ implies $m \wtrans{\sigma} m'$.
Note the difference between $m \wtrans{\epsilon} m'$, which stands for any sequence of $\tau$ transitions,
and $m \ftrans{\epsilon} m'$, which implies $m = m'$ (the sequence is empty).

We use the standard graphical notation for nets, where places are depicted as
circles and transitions as squares such as the nets displayed in
Fig.~\ref{fig:concat}.

\subsection{Polyhedral abstraction}
\label{sec:polyabs}

We define an equivalence relation that can be used to describe a linear
dependence between the markings of two different nets, $N_1$ and $N_2$.
Assume $V$ is a set of places $p_1, \dots, p_n$, considered as variables,
and let $m$ be a mapping in $V \rightarrow \Nat$.
We define $\maseq{m}$ as a linear formula, whose unique model in $\Nat^V$ is $m$,
defined as $\maseq{m} \ \defeq\ \bigwedge \{ x = m(x) \mid x \in V \}$.
By extension, given a Presburger formula $E$, we say that $m$ is a (partial) solution of $E$ if the formula
$E \comma \maseq{m}$ is satisfiable. Equivalently, we can view $\maseq{m}$ as a
substitution, where each variable $x \in V$ is substituted by $m(x)$.
Indeed, the formula $F\{m\}$ (the substitution $\maseq{m}$ applied to $F$)
and $F \land \maseq{m}$ admit the same models.

\subsubsection*{Equivalence Between Markings}

Given two mappings $m_1 \in \Nat^{V_1}$ and $m_2 \in \Nat^{V_2}$, we say
that $m_1$ and $m_2$ are \emph{compatible} when they have equal values on their
shared domain: $m_1(x) = m_2(x)$ for all $x$ in $V_1 \cap V_2$. This is a
necessary and sufficient condition for the system
$\maseq{m_1} \comma \maseq{m_2}$ to be satisfiable.
Finally, if $V$ is the set of free variables of $\maseq{m_1}$, $\maseq{m_2}$,
and the free variables of $E$ are included in $V$,
we say that $m_1$ and $m_2$ are related up-to $E$, denoted $m_1 \eequiv m_2$, when
$E \comma \maseq{m_1} \comma \maseq{m_2}$ is satisfiable.
\begin{equation}
  \label{eq:3}
  m_1 \eequiv m_2  \quad \Leftrightarrow \quad \exists m \in \Nat^V \st m \models E \comma \maseq{m_1} \comma \maseq{m_2}
\end{equation}

\subsubsection*{Equivalence Between Nets}

The previous relation defines an equivalence between markings of two different
nets ($\eequiv \subseteq \Nat^{P_1} \times \Nat^{P_2}$) and, by extension, can
be used to define an equivalence between nets themselves, that is called
\emph{polyhedral equivalence} in~\cite{fi2022,spin2021}, where all reachable
markings of $N_1$ are related to reachable markings of $N_2$ (and conversely),
as explained next.

\begin{definition}[E-Abstraction]
  \label{def:eabs}
  Assume $N_1 \defeq (P_1, T_1, \pre_1, \post_1)$ and
  $N_2 \defeq (P_2, T_2, \pre_2, \post_2)$ are two Petri nets, and $E$ a Presburger formula
  whose free variables are included in $P_1 \cup P_2$.
  We say
  that the marked net $(N_2, m_2)$ is an $E$-abstraction of $(N_1, m_1)$,
  denoted $(N_1, m_1) \eabs_E (N_2, m_2)$, if and only if:
  \begin{description}
  \item[(A1)] the initial markings are related up-to $E$, meaning $m_1 \eequiv
    m_2$;

  \item[(A2)] for all observable sequences $(N_1, m_1) \wtrans{\sigma} (N_1,
    m_1')$ in $N_1$, there is at least one marking $m'_2$ over $P_2$ such that
    $m'_1 \eequiv m'_2$, 
    and for all markings $m_2'$ over $P_2$
    such that $m_1' \eequiv m_2'$ we have $(N_2, m_2) \wtrans{\sigma} (N_2,
    m_2')$.
  \end{description}
  We say that $(N_1, m_1)$ is $E$-equivalent to $(N_2, m_2)$, denoted
  $(N_1, m_1) \reduc_E (N_2, m_2)$, when we have both
  $(N_1, m_1) \eabs_E (N_2, m_2)$ and
  $(N_2, m_2) \eabs_E (N_1, m_1)$.
\end{definition}

By definition, given an equivalence statement $(N_1, m_1) \reduc_E (N_2, m_2)$,
then for every marking $m'_2$ reachable in $N_2$, the set of markings of $N_1$ consistent
with $E \comma \maseq{m'_2}$ is non-empty (condition (A2)).
In practice, this defines a partition of the reachable markings of $(N_1, m_1)$
into a union of ``convex sets''---hence the name polyhedral
abstraction---each associated to one (at least) reachable marking in~$N_2$.

Although $E$-abstraction looks like a simulation, it is not, since the
pair of reachable markings $m'_1, m'_2$ from the definition does not
satisfy $(N_1, m'_1) \eabs_E (N_2, m'_2)$ in general.  This
relation $\eabs_E$ is therefore broader than a simulation, but
suffices for our primary goal, that is Petri net reduction.  Of
course, $\reduc_E$ is not a bisimulation either.

\subsubsection*{Undecidability of the Equivalence Checking}

It is also quite
simple to show that checking $E$-abstraction equivalence is
undecidable in general.

\begin{theorem}[Undecidability of the E-Equivalence Checking]
  \label{th:undecidability-E-equivalence}
  The problem of checking whether a statement $(N_1, m_1) \reduc_E (N_2, m_2)$
  is valid is undecidable.
\end{theorem}

\begin{proof}
  By contradiction, we assume there exists some algorithm, say $\mathcal{A}$,
  that checks the $E$-abstraction equivalence problem.
  More precisely, the input of $\mathcal{A}$ consists in two marked nets $(N_1, m_1)$
  and $(N_2, m_2)$, as well as a Presburger formula $E$ with free variables in the places of $N_1$ and $N_2$.
  The output of $\mathcal{A}$ is a Boolean, indicating whether $(N_1, m_1) \reduc_E (N_2, m_2)$ holds or not.

  Let us consider another problem: given any pair of marked nets $(N_1, m_1)$ and $(N_2, m_2)$ with the same set of places,
  and equal initial markings (i.e., $m_1 = m_2$), check the marking equivalence of both nets, that is check if $R(N_1, m_1) = R(N_2, m_2)$ holds.
  This problem is known to be undecidable*.
  Yet, we will show that algorithm $\mathcal{A}$ is always able to answer to this problem, hence the contradiction.

  Take any pair of marked nets $(N_1, m_1)$ and $(N_2, m_2)$ with the same set of places and $m_1 = m_2$.
  We equip each net with a labeling function $l_1$ (resp. $l_2$) such that $l_1(t) = \tau$
  (resp. $l_2(t) = \tau$) for all transition $t$ of $N_1$ (resp. $N_2$).
  Let us show first that: $(N_1, m_1) \eabs_E (N_2, m_2)$ with the trivial
  constraint $E \defeq \mathrm{True}$ is equivalent to $R(N_1, m_1) \subseteq
  R(N_2, m_2)$.

\medskip
  Condition (A1) trivially holds since $m_1 = m_2$. We now show
  that condition (A2) is necessary and sufficient for $R(N_1, m_1) \subseteq
  R(N_2, m_2)$:
  \begin{itemize}
  \itemsep=0,9pt
    \item Assume condition (A2) holds and take a marking $m_1'$ in $R(N_1,
    m_1)$. We have $m_1' \eequiv m_1'$. Then, by condition (A2) we get $m_1'
    \in R(N_2, m_2)$, and so $R(N_1, m_1) \subseteq R(N_2, m_2)$.

    \item Assume $R(N_1, m_1) \subseteq R(N_2, m_2)$ and take a firing sequence
    $(N_1, m_1) \strans{\varrho_1} (N_1, m_1')$. Since all transitions are
    silent we have $l_1(\varrho_1) = \epsilon$. Both nets share the same sets of
    places, thus $m_1'$ satisfies $m_1' \eequiv m_1'$ (and no other marking
    $m_2'\neq m_1'$ satisfies the condition $m_1' \eequiv m_2'$). By assumption,
    $m_1' \in R(N_2, m_2)$, meaning $(N_2, m_2) \strans{\varrho_2} (N_2, m_1')$
    for some firing sequence $\varrho_2$ such that $l_2(\varrho_2) = \epsilon$,
    and so, condition (A2) holds.
  \end{itemize}

  The statement above is proved. By immediate symmetry, we get that $R(N_1, m_1) = R(N_2, m_2)$
  is equivalent to $(N_1, m_1) \eequiv (N_2, m_2)$.
  As a consequence, checking the marking equivalence problem is equivalent to checking the
  $E$-equivalence problem on $(N_1, m_1)$ and $(N_2, m_2)$, with $E$ the trivial constraint.
  Since algorithm $\mathcal{A}$ is supposed to answer to the latter, it equivalently
  answers to the former, which is a contradiction.

\medskip
  \noindent * Hack proved the undecidability of the marking
  equivalence between two subparts of nets $N_1$, $N_2$ given a pair of initial
  markings not necessary equal~\cite{hack1976decidability}. However, his
  proof's construction leads to the same results when initial markings are
  equal.
\end{proof}

\subsection{Basic properties of polyhedral reduction}
\label{sec:basicreachprop}

We proved in \cite{pn2021,fi2022} that we can use $E$-equivalence to check the
reachable markings of $N_1$ simply by looking at the reachable markings of
$N_2$. We give a first property that is useful when we try to find a
counter-example to a property by looking at firing sequences with increasing
length. Our second property is useful for checking invariants. Both results are
at the basis of our model\linebreak  checker \texttt{SMPT}.

\begin{lemma}[Reachability Checking~\cite{pn2021,fi2022}]
  \label{lemma:reachability}
  Assume $(N_1, m_1) \reduc_E (N_2, m_2)$. Then for all pairs of markings $m_1',
  m'_2$ of $N_1, N_2$ such that $m_1' \eequiv m'_2$ and $m'_2 \in R(N_2,m_2)$ it
  is the case that $m_1' \in R(N_1, m_1)$.
\end{lemma}

\begin{figure}[htbp]
\vspace*{5mm}
  \begin{minipage}[b]{0.47\linewidth}
    \centering
    \includegraphics[width=\textwidth]{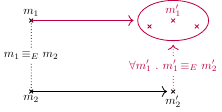}
    \caption[Illustration of the Reachability Checking lemma]{Illustration of Lemma~\ref{lemma:reachability}.}
    \label{fig:reachability}
    \vspace{4ex}
  \end{minipage}
  \hfill
  \begin{minipage}[b]{0.47\linewidth}
    \centering
    \includegraphics[width=\textwidth]{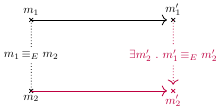}
    \caption[Illustration of the Invariance Checking lemma]{Illustration of Lemma~\ref{lemma:invariance}.}
    \label{fig:invariance}
    \vspace{4ex}
  \end{minipage}\vspace*{-12mm}
\end{figure}

Lemma~\ref{lemma:reachability} (see Fig.~\ref{fig:reachability}) can be used to
find a counter-example $m'_1$ to some property $F$ in $N_1$ (where $F$ is a
formula whose variables are in $P_1$), just by looking at the reachable markings
of $N_2$. Indeed, it is enough to find a marking $m_2'$ reachable in $N_2$ such
that ${m_2'} \models E \wedge \neg F$.

Our second property can be used to prove that every reachable marking of $N_1$
can be traced back to at least one marking of $N_2$ using the reduction
constraints. (While this mapping is surjective, it is not a function since a
state in $N_2$ could be associated with multiple states in $N_1$.)

\begin{lemma}[Invariance Checking~\cite{pn2021,fi2022}]
  \label{lemma:invariance}
  Assume $(N_1, m_1) \reduc_E (N_2, m_2)$. Then for all $m_1'$ in $R(N_1, m_1)$
  there is $m_2'$ in $R(N_2,m_2)$ such that $m_1' \eequiv m_2'$.
\end{lemma}

Using Lemma~\ref{lemma:invariance} (see Fig.~\ref{fig:invariance}), we can
easily extract an invariant on $N_1$ from an invariant on $N_2$. If property $E
\wedge \neg F$ is not reachable on $N_2$, then we can prove that $\neg F$ is not
reachable on $N_1$, meaning $F$ is an invariant. This property (the
\emph{invariant conservation} theorem of Sect.~\ref{sec:smt-based-model})
ensures the soundness of the model checking technique implemented in our tool.

\subsubsection*{Straightforward application}
\label{sec:smt-based-model}

We now recall from \cite{pn2021,fi2022} a general method for combining polyhedral
reductions with SMT-based procedures. Assume we have $(N_1, m_1) \reduc_E (N_2,
m_2)$, where the nets $N_1, N_2$ have sets of places $P_1, P_2$ respectively. In
the following, we use $\vec{p_1} \defeq (p^{1}_{1}, \dotsc, p^{1}_{k})$ and
$\vec{p_2} \defeq (p^{2}_{1}, \dotsc, p^{2}_{l})$ for the places in $P_1$ and
$P_2$. We also consider (disjoint) sequences of variables, $\vec{x}$ and
$\vec{y}$, ranging over (the places of) $N_1$ and $N_2$.
With these notations, we denote $\tilde E(\vec{x}, \vec{y})$ the formula
obtained from $E$ where place names in $N_1$ are replaced with variables in
$\vec{x}$, and place names in $N_2$ are replaced with variables in $\vec{y}$.
When we have the same place in both nets, say $p^{1}_{i} = p^{2}_{j}$, we also
add the constraint $(x_i = y_j)$ to $\tilde{E}$ in order to avoid shadowing
variables. (Remark that $\tilde{E}(\vec{p_1}, \vec{p_2})$ is equivalent to $E$,
since equalities $x_i = y_j$ become tautologies in this case.)
\begin{align}
  \tilde{E}(\vec{x}, \vec{y}) &\defeq E\{\vec{p_1} \leftarrow \vec{x}\} \{\vec{p_2} \leftarrow \vec{y}\}\land \bigwedge_{\{ (i,j) \mid p^{1}_{i} = p^{2}_{j}\}} (x_i = y_j) \label{E_tilde}
\end{align}

Assume $F_1$ is a property that we want to study on $N_1$, such that $\fv{F_1}
\subseteq P_1$. We construct an equivalent formula $F_2$, to study on $N_2$,
which we call the $E$-transform formula of $F_1$.

\begin{definition}[$E$-Transform Formula]
  \label{def:E-transform-formula}
  Assume $(N_1, m_1) \reduc_E (N_2, m_2)$ and take $F_1$ a property with variables in $P_1$, i.e. $\fv{F_1} \subseteq P_1$. Formula $F_2(\vec{y}) \defeq \exists \vec{x} \st \tilde{E}(\vec{x}, \vec{y}) \land F_1(\vec{x})$ is the $E$-transform of $F_1$.
\end{definition}

The following property states that, to check $F_1$ reachable in $N_1$, it is
enough to check the corresponding $E$-transform formula $F_2$ on $N_2$.

\begin{theorem}[Reachability Conservation~\cite{pn2021,fi2022}]
  \label{th:formula_reachability_conservation}
  Assume $(N_1, m_1) \reduc_E (N_2, m_2)$ and that $F_2$ is the $E$-transform of
  formula $F_1$ on $N_1$. Then, formula $F_1$ is reachable in $N_1$ if and only
  if $F_2$ is reachable in $N_2$.
\end{theorem}

Since $F_1$ invariant on $N_1$ is equivalent to $\neg F_1$ not reachable, we can
directly infer an equivalent conservation theorem for invariance:

\begin{corollary}[Invariant Conservation]
  \label{th:invariant_conservation}
  Assume $(N_1, m_1) \reduc_E (N_2, m_2)$ and that $F_2$ is the $E$-transform of
  formula $\neg F_1$ on $N_1$. Then $F_1$ is an invariant on $N_1$ if and only
  if $\neg F_2$ is an invariant on $N_2$.
\end{corollary}

Negating the $E$-transform formula, as done in
Corollary~\ref{th:invariant_conservation}, introduces universally quantified
variables that may impact the solver performance since we require the ``full''
LIA theory instead of only the quantifier-free fragment.
We showed a pragmatic solution to get around this problem in~\cite{amat_project_2024},
where we propose a quantifier elimination procedure specific to the particular
structure of constraints that occur with structural reductions.


\section{Parametric reduction rules and equivalence}
\label{sec:strong_equivalence}

$E$-abstraction is defined on marked nets (Definition~\ref{def:eabs}),
thus the reduction rules defined in~\cite{pn2021,fi2022}, which are $E$-abstraction equivalences,
mention marked nets as well. Their soundness was proven manually, using constrained parameters for initial markings.
Such constraints on markings are called \emph{coherency constraints}.

\subsection{Coherency constraints}

We define a notion of \emph{coherency constraint}, $C$, that must hold
not only in the initial state, but also in a sufficiently large subset
of reachable markings, as formalized next. We have already seen an example with the constraint
$C_1 \defeq (y_2 = 0)$ used in rule \prule{concat}.
%
Without the use of $C_1$, rule \prule{concat} would be unsound since net $N_2$
(right of Fig.~\ref{fig:concat}) could fire transition $b$ more often than its
counterpart,~$N_1$.

Since $C$ is a predicate on markings, we equivalently consider it as a subset of
markings or as a logic formula, so that we may equivalently write $m \models C$
or $m \in C$ to indicate that $C(m)$ is true.
\begin{definition}[Coherent Net]
  \label{def:coherent-net}
  Given a Petri net $N$ and a predicate $C$ on markings, we say that $N$
  satisfies the coherency constraint $C$, or equivalently, that $(N,C)$ is a
  coherent net, if and only if for all firing sequences $m \wtrans{\s} m'$ with
  $m \in C$, we have $$\exists m'' \in C \st  m \ftrans{\s} m'' \wedge m''
  \wtrans{\epsilon} m'$$
\end{definition}

Intuitively, if we consider that all $\tau$ transitions are irreversible
choices, then we can define a partial order on markings with $m < m'$ whenever
$m \trans{\tau} m'$ holds. Then, markings satisfying the coherency constraint
$C$ must be minimal with respect to this partial order.

In this paper, we wish to prove automatically the soundness of a given reduction
rule. A reduction rule basically consists of two nets with their coherency
constraints, and a Presburger relation between markings.
\begin{definition}[Parametric Reduction Rule]
  \label{def:polyhedral_reduction_rule}
  A parametric reduction rule is written $(N_1, C_1) >_E (N_2, C_2)$, where
  $(N_1,C_1)$ and $(N_2,C_2)$ are both coherent nets,
  and $C_1$, $C_2$, and $E$ are Presburger formulas whose free variables are in $P_1 \cup P_2$.
\end{definition}

A given reduction rule $(N_1, C_1) >_E (N_2, C_2)$ is a candidate,
which we will analyze to prove its soundness: is it an $E$-abstraction
equivalence?

\medskip
Our analysis relies on a richer definition of $E$-abstraction, namely parametric $E$-abstraction (Definition~\ref{def:strong}, next),
which includes the coherency constraints $C_1$, $C_2$.
Parametric $E$-abstraction entails $E$-abstraction for each instance of its parameters (Theorem~\ref{th:strong-instance}, below).
Essentially, for any sequence $m_1 \wtrans{\s} m_1'$ with $m_1 \in C_1$, there exists a marking $m_2'$ such that $m_1' \eequiv m_2'$; and
for every marking $m_2 \in C_2$ compatible with $m_1$, i.e., $m_1 \eequiv m_2$,
all markings $m_2'$ compatible with $m'_1$ (i.e., $m_1' \eequiv m_2'$)
can be reached from $m_2$ by the same observable sequence $\s$.
To ease the presentation, we define the notation
\begin{equation}
  \label{eq:cec}
  m_1 \cec m_2  \eqdef m_1 \models C_1 \land m_1 \eequiv m_2 \land m_2 \models C_2
\end{equation}
\begin{definition}[Parametric $E$-Abstraction]
  \label{def:strong}
  Assume $(N_1, C_1) >_E (N_2, C_2)$ is a parametric reduction rule. We say that
  $(N_2, C_2)$ is a parametric $E$-abstraction of $(N_1, C_1)$, denoted $(N_1, C_1)
  \pabs_E (N_2, C_2)$ if and only if:
  \begin{description}
  \item[(S1)] for all markings $m_1$ satisfying $C_1$ there exists a marking $m_2$
  such that $m_1 \cec m_2$;

  \item[(S2)] for all firing sequences $m_1 \wtrans{\epsilon} m'_1$ and all markings $m_2$, we have $m_1 \eequiv m_2$ implies $m'_1 \eequiv m_2$;

  \item[(S3)] for all firing sequences $m_1 \wtrans{\sigma} m_1'$ and all marking pairs $m_2$, $m'_2$,
    if $m_1 \cec m_2$ and $m_1' \eequiv m_2'$ then we have $m_2 \wtrans{\sigma} m_2'$.


  \end{description}
  We say that $(N_1, C_1)$ and $(N_2, C_2)$ are in parametric $E$-equivalence, denoted
  $(N_1, C_1) \preduc_E (N_2, C_2)$, when we have both $(N_1, C_1)
  \pabs_E (N_2, C_2)$ and $(N_2, C_2) \pabs_E (N_1, C_1)$.
\end{definition}

Condition (S1) corresponds to the solvability of the Presburger formula $E$ with
respect to the marking predicates $C_1$ and $C_2$.
Condition (S2) ensures that silent transitions of $N_1$ are abstracted away by the formula $E$,
and are therefore invisible to $N_2$.
Condition (S3) follows closely condition (A2) of the standard $E$-abstraction equivalence.

Note that equivalence $\preduc$ is not a bisimulation, in the same way
that $\reduc$ from Definition~\ref{def:eabs}. It is defined only for
observable sequences starting from states satisfying the coherency
constraints $C_1$ of $N_1$ or $C_2$ of $N_2$, and so this relation is
usually not true on every pair of equivalent markings
$m_1 \eequiv m_2$.

\subsection{Instantiation law}

Parametric $E$-abstraction implies $E$-abstraction for every
instance pair satisfying the coherency constraints $C_1$, $C_2$.

\begin{theorem}[Parametric $E$-Abstraction Instantiation]
  \label{th:strong-instance}
  Assume $(N_1, C_1) \pabs_E (N_2, C_2)$ is a parametric
  $E$-abstraction. Then for every pair of markings $m_1, m_2$ we have
  $m_1 \cec m_2$ implies $(N_1, m_1) \eabs_E (N_2, m_2)$.
\end{theorem}

\begin{proof}
  Consider $(N_1, C_1) \pabs_E (N_2, C_2)$, a parametric $E$-abstraction,
  and $m_1$, $m_2$ such that $m_1 \cec m_2$ holds.
  By definition of $m_1 \cec m_2$ (that is Equation~\eqref{eq:cec}),
  condition (A1) of Definition~\ref{def:eabs} is immediately satisfied.
  We show (A2) by considering an observable sequence $(N_1,m_1) \wtrans{\sigma} (N_1,m'_1)$.
  Since $m_1$ satisfies the coherency constraint $C_1$, we get from Definition~\ref{def:coherent-net}
  a marking $m''_1 \in C_1$ such that $m_1 \ftrans{\sigma} m''_1 \wtrans{\epsilon} m'_1$ holds.
  By applying (S1) to $m''_1$, we get a marking $m'_2$ such that $m''_1 \cec m'_2$ holds, which implies $m''_1 \eequiv m'_2$. Then, by applying
  (S2) to $m''_1 \wtrans{\epsilon} m'_1$, we obtain the expected result $m'_1 \eequiv m'_2$.
  Finally, for all markings $m'_2$ such that $m'_1 \eequiv m'_2$, we conclude $m_2 \wtrans{\sigma} m'_2$ from (S3).
  Condition (A2) is proved, hence $(N_1, m_1) \eabs_E (N_2, m_2)$ holds.
\end{proof}


\section{Automated proof procedure}
\label{sec:procedure}

Our automated proof procedure receives a candidate reduction rule
(Definition~\ref{def:polyhedral_reduction_rule}) as input, and has three
possible outcomes: (i) the candidate is proven sound, congratulations you have
established a new parametric $E$-abstraction equivalence; (ii) the candidate is
proven unsound, try to understand why and fix it (see examples in
Sect~\ref{sec:debugging}); or (iii) we cannot conclude, because part of our
procedure relies on a semi-algorithm for expressing the set of reachable
markings of a flat subnet as a linear constraint (even if the problem is
decidable; see Sects.~\ref{sec:silent_transition_relation}
and~\ref{sec:decidability}).

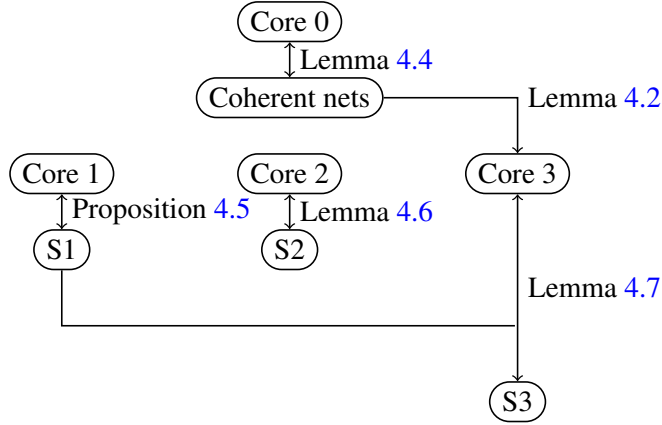
\begin{figure}[tb]
\vspace*{-2mm}
  \centering
\scalebox{0.94}{


\tikzset{cross/.style={cross out, draw=black, fill=none, minimum size=2*(#1-\pgflinewidth), inner sep=0pt, outer sep=0pt}, cross/.default={2pt}}

\begin{tikzpicture}[label distance=6pt*\scalenodes*\scale,x=1pt,y=-1pt,scale=\scale]
	\node[rounded rectangle, draw, semithick] (C0) at (0,0) {Core 0};
	\node[rounded rectangle, draw, semithick] (def0) at (0,50) {Coherent nets};
	\draw[<->, semithick] (C0) -- (def0) node[midway, right] {Lemma~\ref{lem:s0}};

	\node[rounded rectangle, draw, semithick] (C1) at (-150,100) {Core 1};
	\node[rounded rectangle, draw, semithick] (def1) at (-150,150) {S1};
	\draw[<->, semithick] (C1) -- (def1) node[midway, right] {Proposition~\ref{prop:solvable}};

	\node[rounded rectangle, draw, semithick] (C2) at (0,100) {Core 2};
	\node[rounded rectangle, draw, semithick] (def2) at (0,150) {S2};
	\draw[<->, semithick] (C2) -- (def2) node[midway, right] {Lemma~\ref{lem:s2}};

	\node[rounded rectangle, draw, semithick] (C3) at (150,100) {Core 3};
	\node[rounded rectangle, draw, semithick] (def3) at (150,250) {S3};

	\draw[<->, semithick] (C3) -- (def3) node[midway, right] {Lemma~\ref{lem:s3}};
	
	\node (int) at (155,200) {};

	\draw[-, semithick] (def1) |- (int) ;

	\draw[->, semithick] (def0) -| (C3) node[midway, right] {Lemma~\ref{lem:hat}};
\end{tikzpicture} }
  \caption{Detailed dependency relations.}
  \label{fig:detailed_procedure}\vspace*{-4mm}
\end{figure}

\medskip
Given the candidate reduction rule, the procedure generates SMT queries, which we call \emph{core requirements}
(defined in Sect.~\ref{sec:core})
that are solvable if and only if the candidate is a parametric $E$-abstraction
(Theorems~\ref{th:soundness} and \ref{th:completeness}, Sect.~\ref{sec:global-procedure}).
We express these constraints into Presburger predicates, so it is
enough to use solvers for the theory of formulas on Linear Integer
Arithmetic, what is known as LIA in SMT-LIB~\cite{BarFT-RR-17}. We
illustrate the results given in this section using a diagram
(Fig.~\ref{fig:detailed_procedure}) that describe the dependency
relations between conditions (S1), (S2), (S3) and their encoding as
core requirements.

\subsection{Presburger encoding of Petri net semantics}
\label{sec:encoding}

We start by defining a few formulas that
ease the subsequent expression of core requirements. This will help
with the most delicate point of our encoding, which relies on how to encode sequences of transitions.
Note that the coherency constraints of reduction rules are already defined as Presburger formulas.

In the following, we use $\vec{x}$ for the vector of variables $(x_1, \dots, x_n)$,
corresponding to the places $p_1, \dots, p_n$ of $P$, and
$F(\vec{x})$ for a formula whose variables are included in $\vec{x}$. We say that a mapping $m$ of
$\Nat^{P}$ is a \emph{model} of $F$, denoted $m \models F$, if the ground formula
$F(m) \defeq F(m(p_1), \dots, m(p_n))$ is true. Hence, we can also interpret $F$ as a
predicate over markings. Finally, we define the semantics of $F$ as the set
$\sem{F} \defeq \{ m \in \Nat^P \mid m \models F\}$. As usual, we say that a
predicate $F$ is \emph{valid}, denoted $\models F$, when all its interpretations
are true ($\sem{F} = \Nat^P$).
In order to keep track of fired transitions in our encoding, and without any loss of generality we assume that our
alphabet of labels $\Sigma$ is a subset of the natural numbers ($\Sigma \subset \mathbb{N}^*$), except $0$ that
is reserved for $\tau$.

\medskip
We define next a few Presburger formulas that express properties on markings of a net $N$.
For instance, Equation~\eqref{eq:enbl} below defines the predicate $\mathrm{ENBL}_t$,
for a given transition $t$, which corresponds exactly to the markings that enable $t$.
We also define a linear predicate $\mathrm{T}(\vec{x}, \vec{x'}, a)$
that describes the relation between the
markings before ($\vec{x}$) and after ($\vec{x'}$) firing a transition with label $a$.
With this convention, formula $\mathrm{T}(m, m', a)$ holds if and only if
$m \trans{t} m'$ holds for some transition $t$ such that $l(t) = a$ (which implies $a \neq 0$).
\begin{align}
  \mathrm{ENBL}_t(\vec{x}) &\defeq \textstyle \bigwedge_{i \in 1..n} (x_i \geq \pre(t, p_i))\label{eq:enbl}\\
  \Delta_t(\vec{x}, \vec{x}') &\defeq \textstyle  \bigwedge_{i \in 1..n} (x_i' = x_i + \post(t,p_i) - \pre(t, p_i)) \label{eq:delta}\\
  \mathrm{T}(\vec{x}, \vec{x}', a) &\defeq \textstyle \bigvee_{t \in T} \left( \mathrm{ENBL}_t(\vec{x}) \land \Delta_t(\vec{x}, \vec{x}') \land a = l(t) \right)
\end{align}

We admit the following, for all markings $m$, $m'$ and label $a$:
\begin{align}
  \models \mathrm{T}(m, m', a) & \iff \exists t \st m \trans{t} m' \wedge l(t) = a \label{def-t}
\end{align}

In order to define the core requirements, we additionally require a predicate
$\tau_C^*(\vec{x}, \vec{x'})$ encoding the markings reachable by firing any sequence of silent
transitions from a state satisfying the coherency constraints $C$.
And so, the following constraint must hold:

\begin{equation}
  \label{constraint_tau}
  \models m \in C \implies (\tau^*_C(m, m') \iff m \wtrans{\epsilon} m')
\end{equation}

Since $m \wtrans{\epsilon} m'$ may fire an arbitrary number of silent transitions $\tau$, the predicate $\tau_C$ is
not guaranteed to be expressible as a Presburger formula in the general case.
Yet, in Sect.~\ref{sec:silent_transition_relation}, we characterize the Petri nets for which $\tau_C$ can be expressed in Presburger arithmetic,
which include all the polyhedral reductions that we meet in practice (we explain why).

\medskip
Thanks to this predicate, we define the formula $\tleft{\mathrm{T}}_C(\vec{x}, \vec{x'}, a)$ encoding the reachable markings
from a marking satisfying the coherency constraint $C$, by firing any number of silent transitions, followed by a transition labeled with $a$.
Then, we define $\hat{\mathrm{T}}$ which extends $\tleft{\mathrm{T}}$ with any number of silent transitions after $a$
and also allows for only silent transitions (no transition $a$).
\begin{equation}
  \tleft{\mathrm{T}}_C(\vec{x}, \vec{x'}, a) \defeq \exists \vec{y} \st \tau_C^*(\vec{x}, \vec{y}) \land \mathrm{T}(\vec{y}, \vec{x'}, a) \label{tleft}\vspace*{-3mm}
\end{equation}
\begin{align}
  \hat{\mathrm{T}}_C(\vec{x}, \vec{x'}, a) \defeq \ &
  \left(\exists \vec{z} \st \tleft{\mathrm{T}}_C(\vec{x}, \vec{z}, a) \wedge C(\vec{z}) \wedge \tau_C^*(\vec{z},\vec{x'}))\right) \label{hatt}
  \\
  & \vee \left( a = 0 \wedge \tau_C^*(\vec{x},\vec{x'}) \right)
  \label{hatt0}
\end{align}

\begin{lemma}
  \label{lem:tleft}
  For any markings $m,m'$ and label $a$ such that $m \in C$, we have $\models \tleft{\mathrm{T}}_C(m,m',a)$ if and only if $m \ftrans{a} m'$ holds.
\end{lemma}
\begin{proof}
  We show both directions separately.
  \begin{itemize}
  \itemsep=0.9pt
  \item
    Assume $m \ftrans{a} m'$. By definition, this implies that there exists $m''$ and a transition $t$ such that $l(t) = a$
    and $m \wtrans{\epsilon} m'' \trans{t} m'$.
    Therefore, $\tau_C^*(m, m'')$ is valid by Equation~(\ref{constraint_tau}),
    and $\mathrm{T}(m'', m',a)$ is valid by Equation~(\ref{def-t}), hence the expected result $\models \tleft{\mathrm{T}}_C(m,m',a)$.

  \item Conversely, assume $\tleft{\mathrm{T}}_C(m,m',a)$ is valid.
    Then, by Equation~(\ref{tleft}) there exists a marking $m''$ such that
    both $\tau_C^*(m, m'')$ and $T( m'', m',a)$ are valid.
    From Equation~(\ref{constraint_tau}), we get $m \wtrans{\epsilon} m''$,
    and Equation~(\ref{def-t}) implies $\exists t \st m'' \trans{t} m' \wedge l(t) = a$.
    Thus, $m \wtrans{\epsilon} m'' \trans{t} m'$, that is the expected result $m \ftrans{a} m'$.
  \end{itemize}

  \vspace*{-6mm}
\end{proof}

\begin{lemma}
  \label{lem:hat}
  Given a coherent net $(N,C)$, for any markings $m,m'$ such that $m \in C$ and $a \in \Sigma \cup \{0\}$,
  we have $\models \hat{\mathrm{T}}_C(m,m',a)$ if and only if either $m \wtrans{\epsilon} m'$ and $a=0$, or $m \wtrans{a} m'$.
\end{lemma}
\begin{proof}
  We show both directions separately.
  \begin{itemize}
  \item Assume $m \wtrans{\epsilon} m'$ and $a=0$, then $\tau_C^*(m, m')$ is valid by Equation~(\ref{constraint_tau}),
    hence the expected result $\models \hat{\mathrm{T}}_C(m,m',a)$ from Equation~(\ref{hatt0}).

  \item Assume $m \wtrans{a} m'$. From Definition~\ref{def:coherent-net} (coherent net), there exists $m'' \in C$ such that $m \ftrans{a} m'' \wtrans{\epsilon} m'$.
    Then, we get $\models \tleft{\mathrm{T}}_C(m,m'',a)$ from Lemma~\ref{lem:tleft}, and $\models \tau_C^*(m'',m')$ from Equation~(\ref{constraint_tau}).
    Consequently, $\hat{\mathrm{T}}_C(m,m',a)$ is valid from Equation~(\ref{hatt}).

  \item Conversely, assume $\hat{\mathrm{T}}_C(m, m', a)$ holds by Equation~(\ref{hatt0}), then $a=0$ and
    $\models \tau_C^*(m, m')$, which implies $m \wtrans{\epsilon} m'$ by Equation~(\ref{constraint_tau}). This is the expected result.

  \item Finally, assume $\hat{\mathrm{T}}_C(m, m', a)$ holds by Equation~(\ref{hatt}), then there exists a marking $m'' \in C$ such that
    $\models \tleft{\mathrm{T}}_C(m, m'', a)$ and $\models \tau_C^*(m'', m')$. This implies
    $m \ftrans{a} m'' \wtrans{\epsilon} m'$ from Lemma~\ref{lem:tleft} and Equation~(\ref{constraint_tau}).
    This implies the expected result $m \wtrans{a} m'$.
  \end{itemize}

  \vspace*{-7mm}
\end{proof}

As with the $E$-transform formula in Sect.~\ref{def:E-transform-formula}, we denote $\tilde E(\vec{x}, \vec{y})$ the formula
obtained from $E$ where free variables are substituted as follows: place names in $N_1$ are replaced with variables in
$\vec{x}$, and place names in $N_2$ are replaced with variables in
$\vec{y}$ (making sure that bound variables of $E$ are renamed to avoid interference).
When the same place occurs in both nets, say \smash{$p^{1}_{i} = p^{2}_{j}$}, we also add the
equality constraint $(x_i = y_j)$ to $\tilde{E}$ in order to preserve
this equality constraint.

\subsection{Core requirements: parametric $E$-abstraction encoding}
\label{sec:core}

\begin{figure}[!b]
     \centering
    \includegraphics[width=0.4\linewidth]{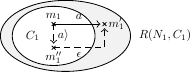}
    \caption{Illustration of (\ref{eq:C0}).}
    \label{fig:def_1}
    \end{figure}

\begin{figure}[ht]
  \begin{minipage}{0.5\textwidth}
    \centering
    \includegraphics[width=0.8\linewidth]{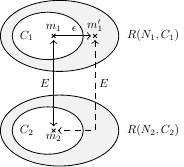}
    \caption{Illustration of (\ref{eq:C2}).}
    \label{fig:def_3}
  \end{minipage}%
  \begin{minipage}{0.5\textwidth}
    \centering
    \includegraphics[width=0.8\linewidth]{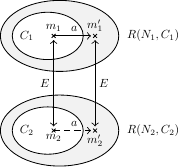}
    \caption{Illustration of (\ref{eq:C3}).}
    \label{fig:def_4}
  \end{minipage}
\end{figure}

In order to check conditions (S1)--(S3) of parametric $E$-abstraction (Definition~\ref{def:strong}),
we define a set of Presburger formulas, called \emph{core requirements}, to be verified using an external SMT solver ((\ref{eq:C1})
to (\ref{eq:C3})). You will find an illustration of these requirements in
Figs.~\ref{fig:def_1}--\ref{fig:def_4}.
The satisfaction of these requirements entail the parametric $E$-abstraction relation.
We have deliberately stressed the notations to prove that $(N_2, C_2)$ is a parametric $E$-abstraction of $(N_1, C_1)$.
Of course, each constraint must be
checked in both directions to obtain the equivalence. Also, in order not to overload the
notations, we assume that the transition relations are clear in the context if
they belong to $N_1$ or $N_2$.

\subsubsection{Verifying that a net is coherent.}
The first step consists in verifying that both nets $N_1$ and $N_2$ satisfy their coherency constraints $C_1$ and $C_2$
(the coherency constraint is depicted in Figure~\ref{fig:def_1}).
We recall Definition~\ref{def:coherent-net}:

\begin{recalldef}{def:coherent-net}{Coherent Net}
  For all firing sequence $m \wtrans{\s} m'$ with $m \in C$, there exists a
  marking $m''$ satisfying $C$ such that $m \ftrans{\s} m''$ and $m''
  \wtrans{\epsilon} m'$.
\end{recalldef}

We encode a simpler relation, below, with sequences $\sigma$ of size 1. This relies on the following result:
\begin{lemma}
  \label{lem:coherent-ftrans}
  $(N,C)$ is coherent if and only if
  for all firing sequence $m \ftrans{a} m'$ with $m \in C$ and $a \in \Sigma$,
  we have $\exists m'' \in C \st  m \ftrans{a} m'' \wedge m'' \wtrans{\epsilon} m'$.
\end{lemma}

We deliberately consider a firing sequence $m \ftrans{a} m'$ (and not $m \wtrans{a} m'$),
since the encoding relies only on $\tleft{\mathrm{T}}_C$ (that is, $\ftrans{a}$), not on $\hat{\mathrm{T}}_C$ (that is, $\wtrans{a}$).

\begin{proof}
  The ``only if'' part is immediate, as a particular case of Definition~\ref{def:coherent-net}
  and noting that $m \ftrans{a} m'$ implies $m \wtrans{a} m'$.
  Conversely, assume the property stated in the lemma is true.
  Then, we show by induction on the size of $\s$, that Definition~\ref{def:coherent-net} holds for any $\s$.
  Note that the base case $\s = \epsilon$ always holds, for any net, by taking $m'' = m$.
  Now, consider a non-empty sequence $\s = \s'.a$ and $m \wtrans{\s'.a} m'$ with $m \in C$.
  By definition, there exists $m_1$ and $m_2$ such that $m \wtrans{\s'} m_1 \ftrans{a} m_2 \wtrans{\epsilon} m'$.
  By induction hypothesis, on $m \wtrans{\s'} m_1$, there exists $m_3 \in C$ such that
  $m \ftrans{\s'} m_3 \wtrans{\epsilon} m_1$.
  Therefore, we have $m \ftrans{\s'} m_3 \wtrans{\epsilon} m_1 \ftrans{a} m_2 \wtrans{\epsilon} m'$, which can simply be written
  $m \ftrans{\s'} m_3 \ftrans{a} m_2 \wtrans{\epsilon} m'$. Using the property stated in the lemma on $m_3 \ftrans{a} m_2$, we get
  a marking $m_4 \in C$ such that $m_3 \ftrans{a} m_4 \wtrans{\epsilon} m_2$.
  Hence, $m \ftrans{\s'} m_3 \ftrans{a} m_4 \wtrans{\epsilon} m_2 \wtrans{\epsilon} m'$ holds, which can be simplified as
  $m \ftrans{\s'.a} m_4 \wtrans{\epsilon} m'$. This is the expected result.
  \end{proof}

Therefore, we can encode Definition~\ref{def:coherent-net}  using the following formula:
\begin{equation}
  \tag{Core 0}
  \begin{split}
    \forall \vec{p}, \vec{p'}, a \st &C(\vec{p}) \land \tleft{\mathrm{T}}_{C}(\vec{p}, \vec{p'}, a)
    \implies \exists \vec{p''} \st C(\vec{p''}) \land \tleft{\mathrm{T}}_{C}(\vec{p}, \vec{p''}, a) \land \tau_{C}^*(\vec{p''}, \vec{p'})
  \end{split}
  \label{eq:C0}
\end{equation}
\begin{lemma}
  \label{lem:s0}
  Given a Petri net $N$, the constraint (\ref{eq:C0}) is valid if and only if the net satisfies the coherency constraint $C$.
\end{lemma}

\begin{proof}
  Constraint (\ref{eq:C0}) is an immediate translation of the property stated in Lemma~\ref{lem:coherent-ftrans}.
\end{proof}

Given a net $N$, a constraint $C$ expressed as a Presburger formula, and a formula $\tau_C^*$ that captures $\wtrans{\epsilon}$ transitions
(as obtained in Sect.~\ref{sec:silent_transition_relation}),
we are now able to check automatically that a net $(N,C)$ is coherent.
Thus, from now on, we assume that the considered nets $(N_1,C_1)$ and $(N_2,C_2)$ are indeed coherent.

\subsubsection{Coherent solvability}
The first requirement of the parametric $E$-abstraction relates to the solvability of formula $E$ with
regard to the coherency constraints $C_1$, and is encoded by \eqref{eq:C1}.
This requirement ensures that every marking of $N_1$ satisfying $C_1$ can be associated to
at least one marking of $N_2$ satisfying $C_2$.
Let us recall (S1), taken from Definition~\ref{def:strong}:
\begin{recalldef}{def:strong}{S1} 
  For all markings $m_1$ satisfying $C_1$ there exists a marking $m_2$ such
  that $m_1 \cec m_2$.
\end{recalldef}

\begin{figure}[!h]
    \centering
    \includegraphics[width=0.4\linewidth]{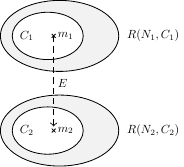}
    \caption{Illustration of (\ref{eq:C1}).}
    \label{fig:def_2}
\end{figure}

Condition (S1) is depicted in Figure~\ref{fig:def_2}.
We propose to encode it by the following Presburger formula:
\begin{equation}
  \tag{Core 1}
  \forall \vec{x} \st C_1(\vec{x}) \implies \exists \vec{y} \st \tilde{E}(\vec{x}, \vec{y}) \land C_2(\vec{y})
  \label{eq:C1}
\end{equation}

Since the encoding is immediate, we admit this proposition:
\begin{proposition}
  \label{prop:solvable}
  The constraint (\ref{eq:C1}) is valid if and only if (S1) holds.
\end{proposition}

\subsubsection{Silent constraints}
So far, we have focused on the specific case of coherent nets, which refers to
intermediate coherent markings. Another notable feature of parametric
$E$-abstractions is the ability to fire any number of silent transitions without
altering the solutions of $E$. In other words, if two markings, $m_1$ and $m_2$,
are solutions of $E$, then firing any silent
sequence from $m_1$ (or $m_2$) will always lead to a solution of $E \land m_2$
(or $E \land m_1$). This means that silent transitions must be invisible to the other net.

Let us recall (S2), taken from Definition~\ref{def:strong}:
\begin{recalldef}{def:strong}{S2} 
  For all firing sequences $m_1 \wtrans{\epsilon} m'_1$ and all markings $m_2$, we have $m_1 \eequiv m_2$ implies $m'_1 \eequiv m_2$.
\end{recalldef}

It actually suffices to show the result for each silent transition $t \in T_1$ taken separately:
\begin{lemma}
  \label{lem:s2}
  Condition (S2) holds if and only if, for all markings $m_1$, $m_2$ such that $m_1 \eequiv m_2$,
  and for all $t_1 \in T_1$ such that $l_1(t_1) = \tau$, we have $m_1 \trans{t_1} m'_1 \imply m'_1 \eequiv m_2$.
  \end{lemma}
\begin{proof}
  The ``only if'' way is only a particular case of (S2) with a single silent transition $t_1$.
  For the ``if'' way, (S2) is shown from the given property by transitivity.
\end{proof}

Thanks to this result, we encode (S2) by the following core requirement:
\begin{equation}
  \tag{Core 2}
\forall \vec{p_1}, \vec{p_2}, \vec{p_1'} \st \tilde{E}(\vec{p_1}, \vec{p_2}) \land \tau(\vec{p_1}, \vec{p_1'}) \implies \tilde{E}(\vec{p_1'}, \vec{p_2})
  \label{eq:C2}
\end{equation}
where $\tau(\vec{x}, \vec{x'})$ is defined as
$\tau(\vec{x}, \vec{x'}) \eqdef \bigvee_{t \in T \mid l(t) = \tau} \left( \mathrm{ENBL}_t(\vec{x}) \land \Delta_t(\vec{x}, \vec{x}')\right)$

\subsubsection{Reachability}

Let us recall the definition of (S3), taken from Definition~\ref{def:strong}:
\begin{recalldef}{def:strong}{S3} 
  For all firing sequences $m_1 \wtrans{\sigma} m_1'$ and all marking pairs $m_2$, $m'_2$,
  if $m_1 \cec m_2$ and $m_1' \eequiv m_2'$ then we have $m_2 \wtrans{\sigma} m_2'$.
\end{recalldef}

Condition (S3) mentions sequences $\sigma$ of arbitrary length.
We encode it with a formula dealing only with sequences of length at most $1$, thanks to the following result:
\begin{lemma}
  \label{lem:s3}
  Given a parametric reduction rule $(N_1, C_1) >_E (N_2, C_2)$ which satisfies condition~(S1),
  then condition~(S3) holds if and only if
  for all firing sequence $m_1 \wtrans{\s} m_1'$ with $\s = \epsilon$ or $\s = a$ with $a \in \Sigma$, and all markings $m_2,m'_2$, we have
  $m_1 \cec m_2 \wedge m_1' \eequiv m_2' \imply m_2 \wtrans{\s} m_2'$.
\end{lemma}

\begin{proof}
  The given property is necessary as a particular case of (S3) taking $\s = a$ or $\s = \epsilon$.
  Conversely, assume the given property holds.
  We show by induction on the size of $\s$ that (S3) holds for any sequence $\s$.
  The base cases $\s = a$ and $\s = \epsilon$ are ensured by hypothesis.
  Now, consider a non-empty sequence $\s = \s'.a$, and $m_1 \wtrans{\s} m'_1$ (i),
  as well as markings $m_2$, $m'_2$ such that $m_1 \cec m_2$ and $m_1' \eequiv m_2'$ holds.
  We have to show $m_2 \wtrans{\sigma} m_2'$.
  From (i), we have $m_1 \wtrans{\s'.a} m'_1$, that is, there exists a marking $u_1$ such that
  $m_1 \wtrans{\s'} u_1 \wtrans{a} m'_1$ (ii).
  By Definition~\ref{def:coherent-net},
  there exists $u'_1 \in C_1$ such that $m_1 \ftrans{\s'} u'_1 \wtrans{\epsilon} u_1$ (iii).
  Also, by condition (S1), there exists a marking $u'_2$ of $N_2$ such that
  $u'_1 \cec u'_2$, which implies $u'_1 \eequiv u'_2$ (iv).
  Hence, by induction hypothesis on $m_1 \wtrans{\sigma'} u'_1$, we have $m_2 \wtrans{\sigma'} u'_2$ ($\alpha$)
  From (iii) and (ii), we get $u'_1 \wtrans{a} m'_1$ (v). Applying the property of the lemma on (iv) and (v),
  we get $u'_2 \wtrans{a} m'_2$ ($\beta$). Combining ($\alpha$) and ($\beta$) leads to $m_2 \wtrans{\sigma'.a} m'_2$,
  that is the expected result $m_2 \wtrans{\sigma} m'_2$.
\end{proof}

Thanks to Lemma~\ref{lem:s3}, we can encode (S3) by the following formula:
\begin{equation}
  \tag{Core 3}
  \begin{split}
    \forall \vec{p_1}, \vec{p_2}, a, \vec{p_1'}, \vec{p_2'} \st &\cec(\vec{p_1}, \vec{p_2}) \land \hat{\mathrm{T}}_{C_1}(\vec{p_1}, \vec{p_1'}) \land \tilde{E}(\vec{p_1'}, \vec{p_2'})
    \implies \hat{\mathrm{T}}_{C_2}(\vec{p_2}, \vec{p_2'})
  \end{split}
  \label{eq:C3}
\end{equation}

\subsection{Global procedure}
\label{sec:global-procedure}

In this section, we consider the full process for proving parametric $E$-abstraction.
We demonstrate that verifying constraints (\ref{eq:C0}) to
(\ref{eq:C3}) is sufficient for obtaining a sound abstraction
(Theorem~\ref{th:soundness}). We also prove that these conditions are necessary (Theorem~\ref{th:completeness}).

\begin{theorem}[Soundness]
  \label{th:soundness}
  Given two nets $N_1, N_2$ and constraints $C_1, C_2$ expressed as Presburger formulas,
  if core requirement (\ref{eq:C0}) holds for both $(N_1,C_1)$ and $(N_2,C_2)$,
  and if core requirements (\ref{eq:C1}), (\ref{eq:C2}), and (\ref{eq:C3}) are valid,
  then the rule is a parametric $E$-abstraction:
  $(N_1, C_1) \pabs_E (N_2, C_2)$.
\end{theorem}
\begin{proof}
  If (\ref{eq:C0}) holds for $(N_1,C_1)$,
  then $(N_1,C_1)$ is a coherent net by Lemma~\ref{lem:s0}.
  Similarly for $(N_2,C_2)$.
  Hence, $(N_1, C_1) >_E (N_2, C_2)$ is a parametric reduction rule.
  By Proposition~\ref{prop:solvable}, and since (\ref{eq:C1}) is valid,
  we get (S1) from Definition~\ref{def:strong}.
  Similarly, by Lemma~\ref{lem:s2}, and since (\ref{eq:C2}) is valid,
  we get (S2).
  Finally, (S3) holds by Lemma~\ref{lem:s3} since (\ref{eq:C3}) is valid
  and since (S1) is known to hold.
  (S1), (S2), (S3) entail $(N_1, C_1) \pabs_E (N_2, C_2)$ by Definition~\ref{def:strong}.
\end{proof}

The converse also holds:
\begin{theorem}[Completeness]
  \label{th:completeness}
  Given a parametric $E$-abstraction $(N_1, C_1) \pabs_E (N_2, C_2)$,
  then core requirements (\ref{eq:C1}), (\ref{eq:C2}), and (\ref{eq:C3}) are valid,
  and (\ref{eq:C0}) holds for both $(N_1,C_1)$ and $(N_2,C_2)$.
\end{theorem}

\begin{proof}
  By hypothesis, conditions (S1), (S2) and (S3) hold and $(N_1,C_1)$ and $(N_2,C_2)$
  are coherent nets.
  Then, Lemma~\ref{lem:s0} implies that (\ref{eq:C0}) holds for both nets.
  Besides, Proposition~\ref{prop:solvable} and Lemmas~\ref{lem:s2} and~\ref{lem:s3}
  ensure that (\ref{eq:C1}), (\ref{eq:C2}), and (\ref{eq:C3}) are valid.
\end{proof}

Consequently, checking $E$-abstraction equivalence amounts to
check that SMT formulas (\ref{eq:C0})-(\ref{eq:C3}) are valid on both nets.

Our approach relies on our ability
to express (arbitrarily long) sequences $m \wtrans{\epsilon} m'$
thanks to a formula~$\tau_C^*(\vec{x},\vec{x'})$. This is addressed in
the next section.


\section{Accelerating the silent transition relation}
\label{sec:silent_transition_relation}

The previous results, including Theorems~\ref{th:soundness} and~\ref{th:completeness},
rely on our ability to express the reachability set of silent transitions as a Presburger predicate, denoted $\tau_C^*$.
Finding a finite formula $\tau_C^*$ that captures an infinite state space
is not granted, since $\tau$-sequences may be of arbitrary length.
However, we now show that, since $\tau$ transitions must be abstracted away by $E$ in order
to define a valid parametric $E$-equivalence (condition (S2)), and since $E$ is itself a Presburger formula,
this implies that $\tau_C^*$ corresponds to the reachability
set of a \emph{flat} subnet~\cite{presburger_vass}, which is expressible as a Presburger formula too.

We define the \emph{silent reachability set} of a net $N$ from a coherent
constraint $C$ as $R_\tau(N, C) \defeq \{m' \mid m \models C \land m
\wtrans{\epsilon} m'\}$.
We now want to find a predicate $\tau_C^*(\vec{p}, \vec{p'})$ that satisfies the relation:
\begin{equation}
  \tag{\ref{constraint_tau}}
  R_\tau(N, C) \defeq \{m' \mid m' \models \exists \vec{x} \st C(\vec{x}) \land \tau^*_C(\vec{x}, \vec{x'})\}
\end{equation}

In order to express the formula $\tau^*_C$, we first use the tool \texttt{FAST}
\cite{bardin2003fast}, designed for the analysis of infinite systems, and that
permits to compute the reachability set of a given Vector Addition System with
States (VASS). Note that a Petri net can be transformed to an equivalent VASS
with the same reachability set, so the formal presentation of VASS can be skipped.
The algorithm implemented in \texttt{FAST} is a semi-procedure, for
which we have some termination guarantees whenever the net is flat~\cite{bardin_fast_2008},
i.e. its corresponding VASS can be unfolded into a VASS
without nested cycles, called a flat VASS. Equivalently, a net $N$ is flat
for some coherent constraint $C$ if its language is flat, that is, there exists
some finite sequence $\varrho_1 \dots \varrho_k \in T^*$ such that for every initial
marking $m \models C$ and reachable marking $m'$ there is a sequence
$\varrho \in \varrho_1^* \dots \varrho_k^*$ such that $m \wtrans{\varrho} m'$.
In short, all reachable markings can be reached by simple sequences, belonging to the language: $\varrho_1^* \dots \varrho_k^*$.
Last but not least, the authors stated in Theorem XI.2 from \cite{presburger_vass} that
a net is flat if and only if its reachability set is Presburger-definable:

\begin{recallthm}{XI.2}{from~\cite{presburger_vass}}
  \label{th:presburger_vass}
  The class of flatable VAS coincides with the class of Presburger VAS.
\end{recallthm}

As a consequence, \texttt{FAST}'s algorithm terminates when its input is Presburger-definable.
We show in Theorem~\ref{th:definable} that given a parametric $E$-abstraction equivalence
$(N_1, C_1) \preduc_E (N_2, C_2)$, the silent reachability sets for both nets
$N_1$ and $N_2$ with their coherency constraints $C_1$ and $C_2$ are
indeed Presburger-definable---we can even provide the expected formulas.
Yet, our computation is complete only if the candidate reduction rule is
a parametric $E$-abstraction equivalence (then, we are able to compute the $\tau^*_C$ relation),
otherwise \texttt{FAST}, and therefore our procedure too, may not terminate.

\begin{theorem}[Silent State Spaces are Presburger-Definable]
  \label{th:definable}
  Given a parametric $E$-abstraction equivalence $(N_1, C_1) \preduc_E (N_2, C_2)$, the
  silent reachability set $R_\tau(N_1, C_1)$ 
  is Presburger-definable.
\end{theorem}

\begin{proof}
  We prove only the result for $(N_1, C_1)$, the proof for $(N_2, C_2)$ is
  similar since $\preduc$ is a symmetric relation.
  We first propose an expression that computes $R_\tau(N_1, m_1)$ for any
  marking $m_1$ satisfying $C_1$. Consider an initial marking $m_1$ in $C_1$.
  From condition (S1) (solvability of $E$), there exists a compatible marking
  $m_2$ satisfying $C_2$, meaning $m_1 \cec m_2$ holds. Take a silent sequence
  $m_1 \wtrans{\epsilon} m_1'$. From condition (S2) (silent stability), we have
  $m_1' \eequiv m_2$. Hence, $R_\tau(N_1, m_1) \subseteq \{m_1' \mid \exists m_2
  \st C_2(m_2) \land \tilde{E}(m_1, m_2) \land \tilde{E}(m_1', m_2)\}$. Conversely, we show
  that all $m'_1$ solution of $\tilde{E}(m'_1, m_2)$ are reachable from $m_1$.
  Take $m_1'$ such that $m_1' \eequiv m_2$. Since we have $m_2 \wtrans{\epsilon}
  m_2$, by condition (S3) we must have $m_1 \wtrans{\epsilon} m_1'$. And finally
  we obtain $R_\tau(N_1, m_1) = \{m_1' \mid m_1' \models \exists \vec{p_1},
  \vec{p_2} \st \underline{m_1}(\vec{p_1}) \land C_2(\vec{p_2}) \land \tilde{E}(\vec{p_1}, \vec{p_2})
  \land \tilde{E}(\vec{p_1'}, \vec{p_2})\}$.

\medskip
  We can generalize this reachability set for all coherent markings satisfying
  $C_1$. We first recall its definition, $R_\tau(N_1, C_1) = \{m_1' \mid \exists
  m_1 \st m_1 \models C_1 \land m_1 \wtrans{\epsilon} m_1'\}$. From condition
  (S1), we can rewrite this set as $\{m_1' \mid \exists m_1, m_2 \st m_1 \cec
  m_2 \land m_1 \wtrans{\epsilon} m_1'\}$ without losing any marking. Finally,
  thanks to the previous result we get $R_\tau(N_1, C_1) = \{m_1' \mid m_1'
  \models P\}$ with $P = \exists \vec{p_1}, \vec{p_2} \st \cec(\vec{p_1},
  \vec{p_2}) \land \tilde{E}(\vec{p_1'}, \vec{p_2})$ a Presburger formula.
  Because of the $E$-abstraction equivalence, (S1) holds in both directions,
  which gives $\forall \vec{p_2} \st C_2(\vec{p_2}) \implies \exists \vec{p_1}
  \st \tilde{E}(\vec{p_1}, \vec{p_2}) \land C_1(\vec{p_1})$. Hence, $P$ can be
  simplified into $\exists \vec{p_2} \st C_2(\vec{p_2}) \land
  \tilde{E}(\vec{p_1'}, \vec{p_2})$.

  Note that this expression of $R_\tau(N,C)$ relies on the fact that the
  equivalence $(N_1, C_1) \preduc_E (N_2, C_2)$ already holds. Thus, we cannot
  conclude that a candidate rule is an $E$-abstraction equivalence using this
  formula at once without the extra validation of \texttt{FAST}.
\end{proof}

\subsubsection*{Verifying FAST Results.}

We have  shown that \texttt{FAST} terminates in case of a correct parametric
$E$-abstraction. We now show that it is possible to check that the
predicates $\tau^*_{C_1}$ and $\tau^*_{C_2}$, computed from the result
of \texttt{FAST} (see Theorem~\ref{th:definable}) are indeed correct.

\medskip
Assume $\tau^*_C$ is, according to \texttt{FAST}, equivalent to the language $\varrho_1^* \dots \varrho_n^*$
with $\varrho_i \in T^*$. We encode this language with the following Presburger predicate
(similar to the one presented in \cite{tacas2022}), which uses
the formulas $H(\sigma^{k_i})$ and $\Delta(\sigma^{k_i})$ defined below:
\begin{equation}
  \tau^*_C(\vec{p^1}, \vec{p^{n + 1}}) \defeq
                                                  \exists k_1 ... k_n, \vec{p^{2}} \dots \vec{p^{n-1}} \st
                                                  \bigwedge_{i
                                                  \in 1 ..n}  \left (
                                                  ( \vec{p^i}
                                                  \geqslant
                                                  H(\varrho_i^{k_i}) )
                                                  \land
                                                  \Delta(\varrho_i^{k_i})(\vec{p^i},
                                                  \vec{p^{i+1}})
                                                   \right )
\end{equation}
This definition introduces acceleration variables $k_i$, encoding the number of
times we fire the sequence $\varrho_i$. The hurdle and delta of the sequence of transitions $\varrho_i^k$, which depends on $k$,
are written $H(\sigma^{k_i})$ and $\Delta(\sigma^{k_i})$, respectively.
Their formulas are given in Equations~\eqref{eq-hurdle-k} and~\eqref{eq-delta-k} below.
Let us explain how we obtain them.

\medskip
First, we define the notion of hurdle $H(\varrho)$ and delta $\Delta(\varrho)$ of an arbitrary sequence $\varrho$,
such that $m \wtrans{\varrho} m'$ holds if and only if (1)
$m \geqslant H(\varrho)$ (the sequence $\varrho$ is fireable), and (2) $m' = m + \Delta(\varrho)$.
This is an extension of the hurdle and delta of a single transition $t$, already used in Formulas~\eqref{eq:enbl} and~\eqref{eq:delta}.
The definition of $H$ and $\Delta$ is inductive:
\begin{align}
  &H(\epsilon) \defeq \vec{0} \text{,}\;  H(t) \defeq \pre(t) \;\text{ and }\; H(\varrho_1 . \varrho_2) \defeq \max \left ( H(\varrho_1), H(\varrho_2) - \Delta(\varrho_1) \right )\label{classical_hurdle}\\
  &\Delta(\epsilon) \defeq \vec{0} \text{,}\; \Delta(t) \defeq \post(t) - \pre(t) \;\text{ and }\; \Delta(\varrho_1 . \varrho_2) \defeq \Delta(\varrho_1) + \Delta(\varrho_2)\label{classical_delta}
\end{align}
where $\max$ is the component-wise max operator.
The careful reader will check by herself that the definitions of $H(\varrho_1 . \varrho_2)$ and $\Delta(\varrho_1 . \varrho_2)$ do not depend on the way the sequence $\varrho_1 . \varrho_2$ is split.

\medskip
From these, we are able to characterize a necessary and sufficient condition
for firing the sequence $\varrho^k$, meaning firing the same sequence $k$ times.
Given $\Delta(\varrho)$, a place $p$ with a negative displacement (say $-d$)
means that $d$ tokens are consumed each time we fire $\varrho$. Hence, we should
budget $d$ tokens in $p$ for each new iteration, and this suffices to enable the $k-1$ more iterations
following the first transition $\varrho$.
Therefore, we have $m \wtrans{\varrho^k} m'$ if and only if (1) $m \models m \geqslant \mathbbm{1}_{> 0}(k) \times
(H(\varrho) + (k-1) \times \max(\vec{0}, - \Delta(\varrho)))$,
with $\mathbbm{1}_{> 0}(k) = 1$ if and only if $k > 0$, and $0$ otherwise,
and (2) $m' = m + k \times \Delta(\varrho)$.
Concerning the token displacement of this sequence $\varrho^k$, it is $k$ times the
one of the non-accelerated sequence $\varrho$.
Equivalently, if we denote by $m^+$ the ``positive''
part of a mapping $\Delta$, such that $\Delta^+(p) \defeq 0$ when $\Delta(p) \leqslant 0$ and $\Delta^+(p)
\defeq \Delta(p)$ when $\Delta(p) > 0$, we get:
\begin{align}
  &H(\varrho^{k}) \defeq \mathbbm{1}_{> 0}(k) \times (H(\varrho) + (k-1) \times \left (- \Delta(\varrho) \right )^+)  \label{eq-hurdle-k} \\
  &\Delta(\varrho^{k}) \defeq k \times \Delta(\varrho) \label{eq-delta-k}
\end{align}

Finally, given a parametric rule $(N_1, C_1) >_E (N_2, C_2)$ we can now check
that the reachability expression $\smash{\tau^*_{C_1}}$ provided by \texttt{FAST}, and encoded as explained above,
corresponds to the solutions of $\exists \vec{p_2} \st \tilde{E}(\vec{p_1, p_2})$ using the
following additional SMT query:
\begin{equation}
  \forall \vec{p_1}, \vec{p_1'} \st C_1(\vec{p_1}) \implies (\exists \vec{p_2} \st \tilde{E}(\vec{p_1}, \vec{p_2}) \land \tilde{E}(\vec{p_1'}, \vec{p_2}) \Longleftrightarrow \tau^*_{C_1}(\vec{p_1}, \vec{p_1'})) \label{eq:fast-eq}
\end{equation}
(and similarly for $\tau^*_{C_2}$).

\medskip
Once the equivalence~(\ref{eq:fast-eq}) above has been validated by a solver,
it is in practice way more efficient
to use the formula $(\exists \vec{p_2} \st \tilde{E}(\vec{p_1}, \vec{p_2}) \land \tilde{E}(\vec{p_1'}, \vec{p_2}))$
inside the core requirements, rather than the formula $\smash{\tau^*_{C_1}}(\vec{p_1}, \vec{p_1'})$ given by \texttt{FAST},
since the latter introduces many new acceleration variables.


\section{Decidability}
\label{sec:decidability}

Even if our method may not terminate, since \texttt{FAST} is only a semi-decision procedure, we
can prove that checking the correctness of parametric $E$-abstraction is
decidable.

\begin{theorem}[Checking Parametric $E$-abstraction is Decidable]
  \label{th:parametric-E-abstraction-decidable}
  Given two nets $N_1, N_2$ and constraints $C_1, C_2$ expressed as Presburger
  formulas. The problem of deciding whether the statement $(N_1, C_1) \preduc_E
  (N_2, C_2)$ holds is decidable.
\end{theorem}

\begin{proof}
  We proved in Theorems~\ref{th:soundness} and~\ref{th:completeness} that the
  statement $(N_1, C_1) \preduc_E (N_2, C_2)$ holds if and only if (\ref{eq:C0})
  is valid for both nets $(N_1, C_1)$ and $(N_2, C_2)$ and core requirements
  (\ref{eq:C1}), (\ref{eq:C2}), and (\ref{eq:C3}) are valid (in both ways).
  Furthermore, checking the truth of Presburger formulas is\linebreak
  decidable~\cite{presburger_completeness_1991}.

  We are left to prove that we can construct these formulas. The crux relies on
  the computation of predicates $\tau_{C_1}^*$ and $\tau_{C_2}^*$. We proved in
  Theorem~\ref{th:definable} a necessary condition to have a correct
  equivalence, that is, $R_\tau(N_1, C_1)$ and $R_\tau(N_2, C_2)$ must be
  Presburger-definable. The problem of deciding if the reachability set of a
  general Petri net from an initial Presburger set of markings is Presburger
  (equivalently semilinear~\cite{ginsburg_semigroups_1966}) is
  decidable~\cite{hauschildt_semilinearity_1990,lambert_vector_1990}. Then, if
  either $R_\tau(N_1, C_1)$ or $R_\tau(N_2, C_2)$ is not Presburger-definable we
  can assert that the equivalence does not hold; without constructing the core
  requirements. Otherwise, the net is flat~\cite{presburger_vass};
  and computing $\tau_{C_1}^*$ and $\tau_{C_2}^*$ is also
  decidable~\cite{finkel_how_2002}.

\medskip
  Hence, we proposed a theoretical procedure to answer the problem of deciding a
  parametric $E$-equivalence holds, where all steps are decidable.
\end{proof}


\section{Generalizing equivalence rules}
\label{sec:construction-laws}

In this section we discuss some results related
with the \emph{genericity} and \emph{generalisability} of our
abstraction rules. We consider several ``dimensions'' in which a rule
can be generalized. A first dimension is related with the
parametricity of the initial marking, which is taken into account by
our use of a parametric equivalence, $\preduc$ instead of $\reduc$,
see Theorem~\ref{th:strong-instance}. Next, we show that we can infer an
infinite number of equivalences from a single abstraction rule using
compositionality, transitivity, and structural modifications involving
labels. Therefore, each abstraction law can be interpreted as a schema
for several equivalence rules.

\begin{definition}[Transition Operations]
  Given a Petri net $N \defeq (P, T, \pre, \post)$ and its labeling
  function $l : T \to \Sigma \cup \{\tau\}$, we define two operations:
  $\tsub$, for removing, and $\tadd$, for duplicating transitions.
  Let $a$ and $b$ be labels in $\Sigma$.
  \begin{itemize}
  \item $\tsub(a)$ is a net $(P, T', \pre', \post')$, where
    $T' \eqdef T \setminus l^{-1}(a)$,
    and $\pre'$ (resp. $\post'$) is the projection of $\pre$ (resp. $\post$) to the domain $T'$.

  \item $\tadd(a,b)$ is a net $(P, T', \pre', \post')$, where $T'$ is a subset of $T \times \{ 0, 1 \}$ defined by
    $T' \eqdef T \times \{0\} \cup l^{-1}(a) \times \{1\}$.
    Additionally, we define $\pre'(t,i) \eqdef \pre(t)$ and $\post'(t,i) \eqdef \post(t)$ for all $t \in T$ and $i \in \{ 0, 1 \}$.
    Finally, the labeling function $l'$ is defined with $l'(t,0) \eqdef l(t)$ and $l'(t,1) = b$ for all $t \in T$.
  \end{itemize}
\end{definition}

The operation $\tsub(a)$ removes transitions labeled by $a$, while
$\tadd(a,b)$ duplicates all transitions labeled by $a$ and labels the
copies with $b$. We illustrated $\tadd$ in the nets of rule
\prule{magic}, in Fig.~\ref{fig:magic}, where the ``dashed''
transition $c'$ can be interpreted has the result of applying
operation $\tadd(c,c')$. Note that these operations only involve
labeled transitions. Silent transitions are kept untouched---up-to
some injection.

\begin{theorem}[Preservation by Transition Operations]
  \label{th:strong-is-preserved}
  Assume we have a parametric $E$-equivalence
  $(N_1, C_1) \preduc_E (N_2, C_2)$, $a$ and $b$ are labels in $\Sigma$. Then,
  \begin{itemize}
  \item $\tsub_i(a)$ and $\tadd_i(a,b)$ satisfy the coherency constraint $C_i$, for $i = 1, 2$.
  \item $(\tsub_1(a), C_1) \preduc_E (\tsub_2(a), C_2)$.
  \item $(\tadd_1(a,b), C_1) \preduc_E (\tadd_2(a,b), C_2)$.
  \end{itemize}
  where $\tsub_i$, $\tadd_i$ is (respectively) the operation $\tsub$, $\tadd$ on $N_i$.
\end{theorem}

\begin{proof}
  We assume $(N_1, C_1) \preduc_E (N_2, C_2)$ (i) holds, which implies
  that $N_1$ satisfies the coherency constraint $C_1$ (resp., $N_2$ satisfies
  $C_2$). For each operation $\tsub$, $\tadd$, we show that the transformed nets
  $N'_1$ and $N'_2$ still satisfy the coherency constraints and that the
  conditions (S1), (S2), (S3) of definition~\ref{def:strong} still hold.
  Conditions (S1) and (S2) do not involve labeled transitions, so they
  immediately hold in $N'_1$ and $N'_2$.
  (S3) is proven by considering each operation separately.
  \begin{itemize}
  \item Case $\tsub(a)$: $N'_1$ (resp. $N'_2$) is $N_1$ (resp. $N_2$) without transitions labeled by $a$.
    Assume $(N'_1,m_1) \wtrans{\sigma} (N'_1,m_1')$ holds (hence, $a \notin
    \sigma$). From (i), for all markings $m_2$, $m'_2$, such that $m_1 \cec m_2
    \wedge m_1' \eequiv m_2'$, we have $(N_2,m_2) \wtrans{\sigma} (N_2,m_2')$.
    Hence, $(N'_2,m_2) \wtrans{\sigma} (N'_2,m_2')$ holds since $a \notin
    \sigma$.

  \item Case $\tadd(a,b)$: $N'_1$ (resp. $N'_2$) is $N_1$ (resp. $N_2$) with
    transitions labeled by $a$ duplicated and duplicates are labeled by $b$.
    Assume $(N'_1,m_1) \wtrans{\sigma} (N'_1,m_1')$ holds. Let $\sigma_a$ be
    $\sigma\subst{b}{a}$. Then, we have $(N_1,m_1) \wtrans{\sigma_a}
    (N_1,m'_1)$. From (i), for all markings $m_2$, $m'_2$, such that $m_1 \cec
    m_2 \wedge m_1' \eequiv m_2'$, we have $(N_2,m_2) \wtrans{\sigma_a}
    (N_2,m_2')$. Then, $(N'_2,m_2) \wtrans{\sigma_a} (N'_2,m_2')$ holds since
    transitions of $N_2$ are included in those of $N'_2$. In $N'_2$, each
    transition labeled by $a$ is identical to a twin transition labeled by $b$.
    Hence, any such transition can be freely replaced by its twin. Therefore,
    $(N'_2,m_2) \wtrans{\sigma} (N'_2,m_2')$ also holds. This concludes the
    case.
  \end{itemize}
  The proof that net $N'_1$ (resp. $N'_2$) still satisfies the coherency
  constraint $C_1$ (resp. $C_2$) is also done by considering each operation
  separately, and is actually very similar to the above cases (we omit the
  details).
  The three conditions (S1), (S2), (S3) hold on $N'_1$ and $N'_2$, thus $(N'_1,
  C_1) \preduc_E (N'_2, C_2)$ is shown.
\end{proof}

Finally, we recall a previous result from \cite{pn2021,fi2022}
(Theorem~\ref{th:congruence}), which states that equivalence rules can
be combined together using synchronous composition, relabeling, and
chaining.
Note that, in order to avoid inconsistencies that could emerge if we
inadvertently reuse the same variable in different reduction equations
(variable escaping its scope), we require that conditions can be
safely composed: the equivalence statements
$(N_1, m_1) \reduc_E (N_2, m_2)$ and
$(N_2, m_2) \reduc_{E'} (N_3, m_3)$ are \emph{compatible} if and only
if $P_1 \cap P_3 = P_2 \cap P_3$.
We also rely on classical operations for relabeling a net, and for
synchronous product, $N_1 \sprod N_2$, which are defined
in~\cite{fi2022} for instance.

\begin{theorem}[$E$-Equivalence is a Congruence~\cite{pn2021,fi2022}]
  \label{th:congruence}
  Assume we have two compatible equivalence statements $(N_1, m_1) \reduc_E
  (N_2, m_2)$ and $(N_2, m_2) \reduc_{E'} (N_3, m_3)$, and that $M$ is a Petri
  net such that $N_1 \sprod M$ and $N_2 \sprod M$ are defined, then
  \begin{itemize}
  \item $(N_1, m_1) \sprod (M, m) \reduc_E (N_2, m_2) \sprod (M, m)$.
  \item $(N_1, m_1) \reduc_{\exists P_2 \setminus (P_1 \cup P_3) . E \land E'} (N_3, m_3)$.
  \item $(N_1[a/b], m_1) \reduc_{E} (N_2[a/b], m_2)$ for any $a \in \Sigma$ and
  $b \in \Sigma \cup \{\tau\}$.
  \end{itemize}
\end{theorem}


\section{Checking the state space partition}
\label{sec:checking-state-space-partition}

We finally propose to check whether a statement provides a state space
partition---that is not entailed by the parametric $E$-equivalence---since the
relation is symmetric---by verifying two additional core requirements
((\ref{eq:C4}) and (\ref{eq:C5})) on the reduced net $N_2$. It is important to
emphasize that the state space partition is not a prerequisite for solving
reachability problems mentioned in Sect.~\ref{sec:basicreachprop}. Nevertheless, it is a
requirement for some model counting methods, for which polyhedral reduction were
initially developed~\cite{berthomieu2018petri,berthomieu_counting_2019}.

Given a marking $m_2'$ of the reduced net $N_2$, we define $\inv_E(m_2')$ as the
set of markings of the initial net $N_1$ related to $m_2'$.
\begin{equation}
  \inv_E(m_2') \defeq \{ m_1' \mid m_1' \eequiv m_2' \}
\end{equation}
\begin{definition}[Equivalence Preserves Partitioning]
  \label{def:partitioning}
  Given a parametric equivalence $(N_1, C_1) \preduc_E (N_2, C_2)$, we say that it preserves partitioning
  if and only if the family of sets $S \defeq \{ \inv_E(m_2') \mid m_2' \in R(N_2, C_2) \}$ is a partition of $R(N_1, C_1)$.
\end{definition}

Note that, although the equivalence $\preduc_E$ is symmetric, the partitioning property is not. That is,
although $(N_1, C_1) \preduc_E (N_2, C_2)$ and $(N_2, C_2) \preduc_E (N_1, C_1)$ both hold,
in general at most one of these relations preserves partitioning.

\medskip
Here are the formulas that we use to check if an equivalence preserves partitioning:
\begin{equation}
  \tag{Core~4}
    \forall \vec{p_2}, \vec{p_2'} \st C_2(\vec{p_2}) \land \tau(\vec{p_2}, \vec{p_2'}) \implies \mathrm{EQ}(\vec{p_2}, \vec{p_2'}) \\
  \label{eq:C4}
\end{equation}
\begin{equation}
  \tag{Core~5}
    \forall \vec{p_1}, \vec{p_2}, \vec{p_2'} \st  C_2(\vec{p_2}) \land C_2(\vec{p_2'}) \land \tilde{E}(\vec{p_1}, \vec{p_2}) \land \tilde{E}(\vec{p_1}, \vec{p_2'}) \implies \mathrm{EQ}(\vec{p_2}, \vec{p_2'})\\
  \label{eq:C5}
\end{equation}

\begin{theorem}[Checking State Space Partition]
  \label{thm:partition}
  The equivalence $(N_1, C_1) \preduc_E (N_2, C_2)$ preserves partitioning
  if and only if (\ref{eq:C4}) and (\ref{eq:C5}) are valid.
\end{theorem}

\begin{proof}
  The set $S$, as defined in Definition~\ref{def:partitioning} is a partition as a consequence of the following points:\\

  \noindent\textbf{No empty set in S.} For any marking $m_2'$ in $R(N_2, C_2)$
  there exists some marking $m_2$ and sequence $\sigma$ such that $m_2 \models
  C_2$ and $m_2 \wtrans{\sigma} m_2'$. By condition (S1) of the parametric
  $E$-abstraction, there is some marking $m_1$ such that $m_1 \cec m_2$. From
  Theorem~\ref{th:strong-instance}, we have $(N_1, m_1) \reduc_E (N_2, m_2)$.
  Now, by condition (A2) of the $E$-abstraction
  (Definition~\ref{def:eabs}), there is some $m_1'$ such
  that $m_1' \eequiv m_2'$. Thus, $\inv_E(m_2')$ is not empty. This implies
  $\emptyset \notin S$.

\smallskip
  \noindent\textbf{The union $\cup_{A \in S} A$ is equal to $R(N_1,C_1)$.}  We
  prove both inclusions separately.

  \begin{itemize}
    \item Take a marking $m_1'$ in $R(N_1, C_1)$. As previously, we still have
    some markings $m_1 \models C_1$ and $m_2 \models C_2$ such that $(N_1, m_1)
    \reduc_E (N_2, m_2)$ and $m_1' \in R(N_1, m_1)$ (by condition (S1) and
    Theorem~\ref{th:strong-instance}). By condition (A2) of the $E$-abstraction,
    there is some marking $m_2'$ such that $m_2' \in R(N_2, m_2)$ and $m_1'
    \eequiv m_2'$. Hence, there is some set $A \in S$ such that $m_1' \in A$ and
    so $R(N_1, C_1) \subseteq \cup_{A \in S} A $.

  \item Now take a set $A$ in $S$ and a marking $m_1' \in A$. By construction,
  there is some marking $m_2'$ in $R(N_2, C_2)$ such that $m_1' \eequiv m_2'$.
  By condition (S1) and Theorem~\ref{th:strong-instance}, there is $(N_1, m_1)
  \reduc_E (N_2, m_2)$ such that $m_1 \models C_1$, $m_2 \models C_2$ and $m_2'
  \in R(N_2, m_2)$. By condition (A2) of
  Definition~\ref{def:eabs} we have $m_1' \in R(N_1, m_1)$.
  Hence, $m_1' \in R(N_1, C_1)$ and so $\cup_{A \in S} A \subseteq R(N_1,
  C_1)$.\smallbreak
  \end{itemize}

  \noindent\textbf{Pairwise disjoint.} Take two different markings $m_2'$ and
  $m_2''$ in $R(N_2, C_2)$. Since $(N_2, C_2)$ is a coherent net, we can find
  some initial and intermediate markings such that $m_2^{(1)} \wtrans{}
  m_2^{(2)} \wtrans{\epsilon} m_2'$ and $m_2^{(3)} \wtrans{} m_2^{(4)}
  \wtrans{\epsilon} m_2''$ with $m_2^{(i)} \models C_2$ for all $i$ in $1..4$.
  And since (\ref{eq:C4}) is valid, we have $m_2' = m_2^{(2)}$ and $m_2'' =
  m_2^{(4)}$ (firing silent transitions from a coherent state do not change the
  marking). Hence, we get $m_2' \models C_2$ (i) and $m_2'' \models C_2$ (ii).

  Now, we prove by contradiction that $\inv_E(m_2') \cap \inv_E(m_2'') = \emptyset$.
  Assume $\inv_E(m_2') \cap \inv_E(m_2'')$ is not empty and take a marking $m_1$
  from it. Hence, $m_1 \eequiv m_2'$ and $m_1 \eequiv m_2''$. From (i) and (ii),
  and the hypothesis $m_2' \not\equiv m_2''$, we contradict the validity of (\ref{eq:C5}).

\medskip
  We are left to prove that the validity of (\ref{eq:C4}) and (\ref{eq:C5}) is a
  necessary condition to obtain such partition. Assume $S$ is a partition of
  $R(N_1, C_1)$.
  \begin{itemize}
   \item Assume that (\ref{eq:C4}) is not valid. Then, there is a pair of
   different markings $m_2, m_2'$ of $N_2$ such that $m_2 \models C_2$ and $m_2
   \wtrans{\epsilon} m_2'$. From condition (S1) there is some marking $m_1$ such
   that $m_1 \eequiv m_2$, and by condition (S2) we also have $m_1 \eequiv
   m_2'$. Then, there are two sets $A$ and $A'$ in $S$ such that $m_1 \in A$ and
   $m_1 \in A'$, which contradicts that sets in $S$ are pairwise disjoint.

 \item Assume that (\ref{eq:C5}) is not valid. Then, there are some markings
 $m_1$ of $N_1$ and $m_2, m_2'$ of $N_2$ such that $m_2 \models C_2$, $m_2'
 \models C_2$, $m_1 \eequiv m_2$, $m_1 \eequiv m_2'$ and $m_2 \not\equiv m_2'$.
 By construction of $S$, we can find some sets $A, A'$ in $S$, such that $m_1
 \in A$ and $m_1 \in A'$, which also contradicts that sets in $S$ are pairwise
 disjoint.
  \end{itemize}

  \vspace*{-6mm}
\end{proof}

From the proof of Theorem~\ref{thm:partition} we can derive an interesting
characterization on the partitioning of parametric equivalences:

\begin{lemma}
  \label{lem:partitioning}
  A parametric equivalence $(N_1, C_1) \preduc_E (N_2, C_2)$ preserves partitioning if and only if
  for all $m_2 \in R(N_2,C_2)$ and for all $m_1, m'_2$ in $\Nat^{P_1} \times \Nat^{P_2}$,
  $$ m_1 \eequiv m_2 \wedge m_1 \eequiv m'_2 \imply m_2 = m'_2$$
\end{lemma}
\begin{proof}
  Assume the equivalence $(N_1, C_1) \preduc_E (N_2, C_2)$ preserves partitioning.
  Take $m_2 \in R(N_2,C_2)$ and $m_1, m_2$ in $\Nat^{P_1} \times \Nat^{P_2}$, such that
  $m_1 \eequiv m_2 \wedge m_1 \eequiv m'_2$.
  From $m_1 \eequiv m_2$ and $m_2 \in R(N_2,C_2)$, we get $m_1 \in R(N_1,C_1)$  as a consequence of Definition~\ref{def:strong}.
  Then, similarly from $m_1 \eequiv m'_2$, we get $m'_2 \in R(N_2,C_2)$.
  Hence, $m_1 \in \inv_E(m_2)$ and $m_1 \in \inv_E(m'_2)$. Since $S$ (from Definition~\ref{def:partitioning}) is a partition,
  $m_1$ can only belong to exactly one of the sets that constitute $S$. Hence, necessarily $m_2 = m'_2$ (otherwise, two sets of $S$ would contain $m_1$).

\medskip
  Conversely, let us assume that for all $m_2 \in R(N_2,C_2)$ and for all $m_1, m_2$ in $\Nat^{P_1} \times \Nat^{P_2}$,
  $ m_1 \eequiv m_2 \wedge m_1 \eequiv m'_2 \imply m_2 = m'_2$ (i).
  Let $S \defeq \{ \inv_E(m_2') \mid m_2' \in R(N_2, C_2) \}$. We have to show that $S$ is a partition of $R(N_1, C_1)$.
  \smallbreak

  \noindent\textbf{No empty set in S.} Same proof than in Theorem~\ref{thm:partition}.
  \smallbreak

  \noindent\textbf{The union $\cup_{A \in S} A$ is equal to $R(N_1,C_1)$.} Same proof than in Theorem~\ref{thm:partition}.
  \smallbreak

  \noindent\textbf{Pairwise disjoint.} Take two different markings $m_2$ and $m'_2$ in $R(N_2,C_2)$.
  We get the expected result $\inv_E(m_2) \cap \inv_E(m'_2) = \emptyset$ as an immediate consequence of implication (i).

  Consequently, $S$ is a partition of $R(N_1,C_1)$.
  \end{proof}

Once an equivalence is shown to preserve partitioning, it can be freely composed with other
partitioning equivalences. More precisely,

\begin{lemma}
  If the equivalences $(N_1, C_1) \preduc_E (N_2, C_2)$ and $(N_2, C_2) \preduc_{E'} (N_3, C_3)$ are compatible and both preserve partitioning,
  then
  $(N_1, C_1) \preduc_{\exists P_2 \setminus (P_1 \cup P_3) \, E \land E'} (N_3, C_3)$ holds and preserves partitioning too.
\end{lemma}
\begin{proof}
  We assume both equivalences  $(N_1, C_1) \preduc_E (N_2, C_2)$ and $(N_2, C_2)
  \preduc_{E'} (N_3, C_3)$ are compatible and preserve partitioning. Let $E''$
  be ${\exists P_2 \setminus (P_1 \cup P_3) \, E \land E'}$. By virtue of
  Theorem~\ref{th:congruence}, $(N_1, C_1) \preduc_{E''} (N_3, C_3)$ holds. It
  remains to show that it preserves partitioning. Following
  Lemma~\ref{lem:partitioning}, let us take $m_3 \in R(N_3,C_3)$ and $m_1,m'_3
  \in \Nat^{P_1} \times \Nat^{P_3}$, and assume $m_1 \eequiv[E''] m_3$ and $m_1
  \eequiv[E''] m'_3$ both hold. Notice that $m_1 \in R(N_1,C_1)$ and $m'_3 \in
  R(N_3,C_3)$ (as already shown in the proof of Lemma~\ref{lem:partitioning}).
  By definition of $m_1 \eequiv[E''] m_3$, and by removing the existential, we
  get $\exists m \in \Nat^{P_1 \cup P_2 \cup P_3} \st m \models E \land E' \land
  \maseq{m_1} \land \maseq{m_3}$. By projecting $m$ on $P_2$, we get a marking $m_2 \in
  \Nat^{P_2}$ such that $m_1 \eequiv[E] m_2$ and $m_2 \eequiv[E'] m_3$.
  Similarly, there exists $m'_2 \in \Nat^{P_2}$ such that $m_1 \eequiv[E] m'_2$
  and $m'_2 \eequiv[E'] m'_3$. Since $(N_1, C_1) \preduc_E (N_2, C_2)$ preserves
  partitioning, and by Lemma~\ref{lem:partitioning}, we get $m_2 = m'_2$. Then
  again, since $(N_2, C_2) \preduc_{E'} (N_3, C_3)$ preserves partitioning, and
  by Lemma~\ref{lem:partitioning}, we get $m_3 = m'_3$, which is the expected
  result.
  \end{proof}


\section{Debugging reduction rules}
\label{sec:debugging}

An interesting feature of \texttt{Reductron} is to return which core requirements
failed when a rule is unsound. It allows us to pinpoint the problematic condition
and, if necessary, fix it. In this section, we give concrete examples of how this
information can be used to debug a problematic equivalence rule. All our examples
are obtained by mutating one of the
reduction rules used in our toolchain.

\subsubsection*{Modified Equivalence Rule \prule{concat}.}

As we already mentioned, the equivalence rule \prule{concat} becomes incorrect
by adding the ``red dashed'' transition with a label $d$. Our method will output
that $(N_1, C_1)$ is not a coherent net (i.e., (\ref{eq:C0}) fail). In fact,
from a coherent state $m_1 \models C_1$, meaning $m_1(y_2) = 0$, it is not
possible to fire the transition $d$ and then reach a coherent state. This
rule can be fixed by adding a $\tau$ transition from $y_2$ to $y_1$. This new
rule is called \prule{agg} in our reduction framework.

An equivalent incorrect version of the rule \prule{concat}, is the one in
Fig.\ref{fig:concat}, without transitions $d$, but with the coherency constraint
$C_1 \defeq \mathrm{True}$. Our tool, asserts that the expression $\tau_{C_1}^*$
returned by \texttt{FAST} does not correspond to what is expected to obtain a
correct equivalence (see Theorem~\ref{th:definable}). In fact, the tokens
initially contained in $y_2$ cannot be transferred to $y_1$. As previously, one possible
modification to ensure the equivalence is correct when $C_1 \defeq \mathrm{True}$
is to add a silent transition from $y_2$ to $y_1$.

\subsubsection*{Modified Equivalence Rule \prule{magic}}

Now we consider an incorrect version of the \prule{magic} rule depicted in
Fig.~\ref{fig:magic}, with $E \defeq x = y_1 + y_2 + y_3$ (we forget $y_4$). Of
course, again the preliminary check of the predicate $\tau_{C_1}^*$ obtained
using \texttt{FAST} result fail. But, we also obtain that the requirement
(\ref{eq:C2}) does not hold. A counter-example is $m_2 \defeq (x = 1)$, $m_1
\defeq (y_2 = 1) \land (y_1 + y_3 + y_4 = 0)$ and $m_1' \defeq (y_4' = 1) \land
(y_1' + y_2' + y_3' = 0)$. This indicates that there is a problem with the definition
of $E$.

\section{Experimental validation}
\label{sec:examples}

We have implemented our automated procedure in a new tool called
\texttt{Reductron}. The tool is open-source, under the GPLv3 license,
and is freely available on GitHub~\cite{reductron}. The repository
contains a subdirectory, \texttt{rules}, that provides examples of
equivalence rules that can be checked using our approach. Each test
contains two Petri nets, one for $N_1$ (called \texttt{initial.net})
and another for $N_2$ (called \texttt{reduced.net}), defined using the
syntax of \Tina~\cite{laas-cnrs_file_2020}. These nets also include declarations for constraints,
$C_1$ and $C_2$, and for the equation system $E$. Our list contains
examples of laws that are implemented in \texttt{Tedd} and
\texttt{SMPT}, such as rule \prule{concat} depicted in
Fig.~\ref{fig:concat}, but also some examples of unsound equivalences
rules. For instance, we provide example \prule{fake\_concat}, which
corresponds to the example of Fig.~\ref{fig:concat} with transition
$d$ added.

\begin{table}[!h]
  \centering
  \caption{Computation times (time in seconds).\label{tab}}\vspace*{-6mm}
\begin{center}
  \begin {tabular}{llrr}%
  \hline
      Rule   \hspace{5mm} &  \texttt{FAST}  \hspace{5mm} &  \  \texttt{z3}  \hspace{5mm} &   Total \\
\hline
    \prule{agg}                   &  0.09  &  0.18 &  0.28 \\
    \prule{buffer}               & 0.45   & 0.44  & 0.89 \\
    \prule{concat}              & 0.08   & 0.14  & 0.23 \\
    \prule{magic}               & 0.17   & 0.32   & 0.51 \\
    \prule{red}                   &  0.00  & 0.14 &  0.14 \\
    \prule{sink}                  &  0.08  &  0.14 & 0.23 \\
    \prule{SwimmingPool} &  1.92   & 12.11  & 14.03 \\
  \hline
    \end{tabular}
\end{center}\vspace*{-2mm}
\end{table}

We performed some experimentation using \texttt{z3}~\cite{de2008z3} (version
4.8) as our target SMT solver, and \texttt{FAST} (version 2.1). We display the
computation times obtained on all our examples of sound rules in
Table~\ref{tab}. In each case, we give the total time, and how much time was
spent in the two main steps of the procedure, for \texttt{FAST} and \texttt{z3}.
Our results show that our reduction rules can all be checked in a few
seconds.

\newcolumntype{x}[1]{>{\centering\arraybackslash\hspace{0pt}}p{#1}}

\begin{figure}[ht]
  \centering
  \includegraphics[width=0.91\textwidth]{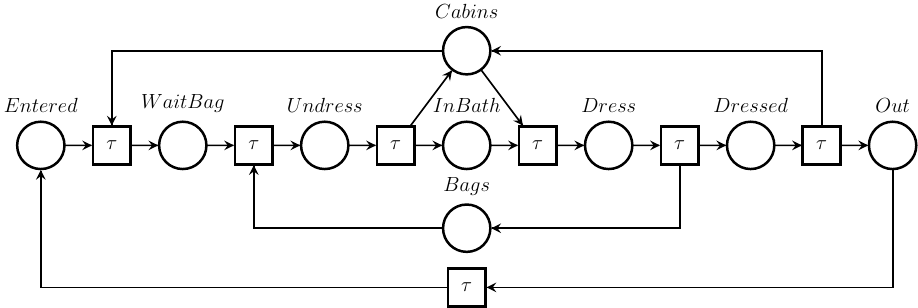}
  $N_1$
  \vskip 2ex
  $C_1 \defeq Cabins = 10 \land Out = 20 \land Bags = 15 \ \land$ \\ $Entered + WaitingBag + Undress + Dresse + Inbath + Dressed = 0$
  $$
  E \defeq \left\{
      \begin{array}{ll}
          Cabins + Dress + Dressed + Undress + WaitBag = 10\\
          Dress + Dressed + Entered + InBath + Out + Undress + WaitBag = 20\\
          Bags + Dress + InBath + Undress = 15
      \end{array}
  \right.
  $$ \vspace*{-3mm}
  \caption{A Petri net modeling users in a swimming pool, see e.g.~\cite{10.1007/3-540-48320-9_14}.\label{fig:Swimming_pool}}\vspace*{-2mm}
\end{figure}

\smallskip
Although we focus on the automatic verification of abstraction laws,
we have also tested our tool on moderate-sized nets, such as the
swimming pool example given in Fig.~\ref{fig:Swimming_pool}. In this
context, we use the fact that an equivalence of the form
$(N, C) \preduc_E (\emptyset, \mathrm{True})$, between $N$ and a net
containing an empty set of places, entails that the reachability set
of $(N, C)$ must be equal to the solution set of $E$. In this case,
also, results are almost immediate.

\medskip
These very good results depend largely on the continuous improvements
made by SMT solvers. Indeed, we generate very large LIA formulas, with
sometimes hundreds of quantified variables, and a moderate amount of
quantifier alternation (formulas of the form
$\forall\, \exists\, \forall$). For instance, experiments performed
with older versions of \texttt{z3} (such as 4.4.1, October 2015)
exhibit significantly degraded performances. We also rely on the very
good performances exhibited by the tool \texttt{FAST}, which is
essential in the implementation of \texttt{Reductron}.

\section{Conclusion}

This work aims to improve the safety of our polyhedral reduction framework using
automated reasoning techniques. But the result we find the most interesting is
the fact that it enhances our understanding of the theoretical underpinnings of
polyhedral equivalence and its close relationship with the notion of flat nets.
It also underlines the importance of coherency constraints, which takes a
central role in our definition of a parametric version of polyhedral
equivalence.
We also hope that it helps better understand how to construct new reduction
rules in the future.

\medskip
There is still ample room to study polyhedral reduction. For instance, we are
interested in characterizing Petri nets that are \emph{fully reducible}, but
where $E$ is a ``convex'' predicate (to ensure that the equivalence defines a
partition of the state space). This defines an interesting and non-trivial
subset of flat nets.

Finally, we exhibited a concrete use case for the problem of deciding whether
the state space of a given Petri net is Presburger-definable. This result can be
found in two different
works~\cite{hauschildt_semilinearity_1990,lambert_vector_1990}, with proofs that
do not easily translate into practical algorithms. We believe that it would be
worthwhile to revisit this problem.

\subsubsection*{Acknowledgements} We would like to thanks J\'er\^ome
Leroux for his support during our experimentation with \texttt{FAST}.



\begin{thebibliography}{10}
\providecommand{\url}[1]{\texttt{#1}}
\providecommand{\urlprefix}{URL }
\expandafter\ifx\csname urlstyle\endcsname\relax
  \providecommand{\doi}[1]{doi:\discretionary{}{}{}#1}\else
  \providecommand{\doi}{doi:\discretionary{}{}{}\begingroup
  \urlstyle{rm}\Url}\fi
\providecommand{\eprint}[2][]{\url{#2}}

\bibitem{berthomieu2018petri}
Berthomieu B, Le~Botlan D, Dal~Zilio S.
\newblock Petri net {Reductions} for {Counting} {Markings}.
\newblock In: Model {Checking} {Software} ({SPIN}), volume 10869 of
  \emph{Lecture Notes in Computer Science}. Springer, 2018
  \doi{10.1007/978-3-319-94111-0_4}.

\bibitem{berthomieu_counting_2019}
Berthomieu B, Le~Botlan D, Dal~Zilio S.
\newblock Counting {Petri} net markings from reduction equations.
\newblock \emph{International Journal on Software Tools for Technology
  Transfer}, 2019.
\newblock \textbf{22}(2):163--181.
\newblock \doi{10.1007/s10009-019-00519-1}.

\bibitem{berthelot_checking_1985}
Berthelot G, Lri-Iie.
\newblock Checking properties of nets using transformations.
\newblock In: Advances in Petri Nets (APN), volume 222 of \emph{Lecture Notes
  in Computer Science}. Springer, 1985 \doi{10.1007/BFb0016204}.

\bibitem{berthelot_transformations_1987}
Berthelot G.
\newblock Transformations and {Decompositions} of {Nets}.
\newblock In: Petri {Nets}: {Central} {Models} and {Their} {Properties} (ACPN),
  volume 254 of \emph{Lecture Notes in Computer Science}. Springer, 1987
 doi:10.1007/ 978-3-540-47919-2\_13.

\bibitem{fi2022}
Amat N, Berthomieu B, Dal~Zilio S.
\newblock A {Polyhedral} {Abstraction} for {Petri} {Nets} and its {Application}
  to {SMT}-{Based} {Model} {Checking}.
\newblock \emph{Fundamenta Informaticae}, 2022.
\newblock \textbf{187}(2-4):103--138.
\newblock \doi{10.3233/FI-222134}.

\bibitem{sttt2022}
Amat N, Dal~Zilio S, Le~Botlan D.
\newblock Leveraging polyhedral reductions for solving {Petri} net reachability
  problems.
\newblock \emph{International Journal on Software Tools for Technology
  Transfer}, 2023.
\newblock \textbf{25}(1):95--114.
\newblock \doi{10.1007/s10009-022-00694-8}.

\bibitem{besson_polyhedral_1999}
Besson F, Jensen T, Talpin JP.
\newblock Polyhedral Analysis for Synchronous Languages.
\newblock In: Static {Analysis} ({SAS}), volume 1694 of \emph{Lecture Notes in
  Computer Science}. Springer, 1999 \doi{10.1007/3-540-48294-6_4}.

\bibitem{feautrier_automatic_1996}
Feautrier P.
\newblock Automatic parallelization in the polytope model.
\newblock In: The {Data} {Parallel} {Programming} {Model}, volume 1132 of
  \emph{Lecture Notes in Computer Science}. Springer, 1996.
\newblock \doi{10.1007/3-540-61736-1_44}.

\bibitem{thierry2009hierarchical}
Thierry-Mieg Y, Poitrenaud D, Hamez A, Kordon F.
\newblock Hierarchical Set Decision Diagrams and Regular Models.
\newblock In: Tools and {Algorithms} for the {Construction} and {Analysis} of
  {Systems} ({TACAS}), volume 5505 of \emph{Lecture Notes in Computer Science}.
  Springer, 2009 \doi{10.1007/978-3-642-00768-2_1}.

\bibitem{tinaToolbox}
{LAAS-CNRS}.
\newblock Tina {Toolbox}, 2023.
\newblock \urlprefix\url{http://projects.laas.fr/tina}.

\bibitem{smpt}
Amat N.
\newblock {SMPT}: {The} {Satisfiability} {Modulo} {Petri} {Nets} {Model}
  {Checker}. {An} {SMT}-based model checker for {Petri} nets focused on
  reachability problems that takes advantage of polyhedral reduction., 2020.
\newblock \urlprefix\url{https://github.com/nicolasAmat/SMPT}.

\bibitem{fm2023}
Amat N, Dal~Zilio S.
\newblock {SMPT}: {A} {Testbed} for {Reachabilty} {Methods} in {Generalized}
  {Petri} {Nets}.
\newblock In: Formal Methods (FM), volume 14000 of \emph{Lecture Notes in
  Computer Science}. Springer, 2023 \doi{10.1007/978-3-031-27481-7_25}.

\bibitem{tacas2022}
Amat N, Dal~Zilio S, Hujsa T.
\newblock Property {Directed} {Reachability} for {Generalized} {Petri} {Nets}.
\newblock In: Tools and {Algorithms} for the {Construction} and {Analysis} of
  {Systems} ({TACAS}), volume 13243 of \emph{Lecture Notes in Computer
  Science}. Springer, 2022 \doi{10.1007/978-3-030-99524-9_28}.

\bibitem{mcc2019}
Amparore E, Berthomieu B, Ciardo G, Dal~Zilio S, Gall{\`a} F, Hillah LM,
  Hulin-Hubard F, Jensen PG, Jezequel L, Kordon F, Le~Botlan D, Liebke T,
  Meijer J, Miner A, Paviot-Adet E, Srba J, Thierry-Mieg Y, van Dijk T, Wolf K.
\newblock Presentation of the 9th Edition of the Model Checking Contest.
\newblock In: Tools and Algorithms for the Construction and Analysis of Systems
  ({TACAS}), {LNCS}. Springer, 2019 \doi{10.1007/978-3-662-58381-4_9}.

\bibitem{esparza1994decidability}
Esparza J, Nielsen M.
\newblock Decidability issues for {Petri} nets.
\newblock \emph{BRICS Report Series}, 1994.
\newblock \textbf{1}(8).
\newblock \doi{10.7146/brics.v1i8.21662}.

\bibitem{esparza1998decidability}
Esparza J.
\newblock Decidability and complexity of {Petri} net problems --- {A}n
  introduction.
\newblock In: Lectures on {Petri} {Nets} {I}: {Basic} {Models} ({ACPN}), volume
  1491 of \emph{Lecture Notes in Computer Science}. Springer, 1998
  \doi{10.1007/3-540-65306-6_20}.

\bibitem{hack1976decidability}
Hack MHT.
\newblock Decidability Questions for {Petri} {Nets}.
\newblock {PhD} {Thesis}, 1976.

\bibitem{hirshfeld1994petri}
Hirshfeld Y.
\newblock Petri nets and the equivalence problem.
\newblock In: {Computer} {Science} {Logic} ({CSL}), volume 832 of \emph{Lecture
  Notes in Computer Science}. Springer, 1994 \doi{10.1007/BFb0049331}.

\bibitem{pn2021}
Amat N, Berthomieu B, Dal~Zilio S.
\newblock On the {Combination} of {Polyhedral} {Abstraction} and {SMT}-{Based}
  {Model} {Checking} for {Petri} {Nets}.
\newblock In: Application and {Theory} of {Petri} {Nets} and {Concurrency}
  ({PETRI} {NETS}), volume 12734 of \emph{Lecture Notes in Computer Science}.
  Springer, 2021 \doi{10.1007/978-3-030-76983-3_9}.

\bibitem{DBLP:journals/corr/abs-2006-05600}
Hujsa T, Berthomieu B, {Dal Zilio} S, {Le Botlan} D.
\newblock Checking marking reachability with the state equation in {Petri} net
  subclasses.
\newblock \emph{CoRR}, 2020.
\newblock \textbf{abs/2006.05600}.

\bibitem{DBLP:journals/corr/abs-2005-04818}
Hujsa T, Berthomieu B, {Dal Zilio} S, {Le Botlan} D.
\newblock On the {Petri} Nets with a Single Shared Place and Beyond.
\newblock \emph{CoRR}, 2020.
\newblock \textbf{abs/2005.04818}.
\newblock \eprint{2005.04818}.

\bibitem{10.1007/3-540-48320-9_14}
B\'erard B, Fribourg L.
\newblock Reachability {Analysis} of ({Timed}) {Petri} {Nets} {Using} {Real}
  {Arithmetic}.
\newblock In: Concurrency {Theory} ({CONCUR}), volume 1664 of \emph{Lecture
  Notes in Computer Science}. Springer, 1999 \doi{10.1007/3-540-48320-9_14}.

\bibitem{bardin2003fast}
Bardin S, Finkel A, Leroux J, Petrucci L.
\newblock {FAST}: {Fast} Acceleration of Symbolic Transition Systems.
\newblock In: Computer {Aided} {Verification} ({CAV}), volume 2725 of
  \emph{Lecture Notes in Computer Science}. Springer, 2003
  \doi{10.1007/978-3-540-45069-6_12}.

\bibitem{bardin_fast_2008}
Bardin S, Finkel A, Leroux J, Petrucci L.
\newblock {FAST}: acceleration from theory to practice.
\newblock \emph{International Journal on Software Tools for Technology
  Transfer}, 2008.
\newblock \textbf{10}(5):401--424.
\newblock \doi{10.1007/s10009-008-0064-3}.

\bibitem{presburger_vass}
Leroux J.
\newblock Presburger {Vector} {Addition} {Systems}.
\newblock In: {Logic} in {Computer} {Science} (LICS). IEEE, 2013
  \doi{10.1109/LICS.2013.7}.

\bibitem{amat_automated_2023}
Amat N, Dal~Zilio S, Le~Botlan D.
\newblock Automated {Polyhedral} {Abstraction} {Proving}.
\newblock In: Application and {Theory} of {Petri} {Nets} and {Concurrency}
  ({PETRI} {NETS}), volume 13929 of \emph{Lecture Notes in Computer Science}.
  Springer, 2023 \doi{10.1007/978-3-031-33620-1_18}.

\bibitem{spin2021}
Amat N, Dal~Zilio S, Le~Botlan D.
\newblock Accelerating the {Computation} of {Dead} and {Concurrent} {Places}
  {Using} {Reductions}.
\newblock In: Model {Checking} {Software} ({SPIN}), volume 12864 of
  \emph{Lecture Notes in Computer Science}. Springer, 2021
  \doi{10.1007/978-3-030-84629-9_3}.

\bibitem{amat_project_2024}
Amat N, Dal~Zilio S, Le~Botlan D.
\newblock Project and {Conquer}: {Fast} {Quantifier} {Elimination} for
  {Checking} {Petri} {Nets} {Reachability}.
\newblock In: Verification, Model Checking, and Abstract Interpretation
  (VMCAI), Lecture Notes in Computer Science. Springer, 2024
  \doi{10.1007/978-3-031-50524-9_5}.

\bibitem{BarFT-RR-17}
Barrett C, Fontaine P, Tinelli C.
\newblock The {SMT}-{LIB} {Standard}: {Version} 2.6.
\newblock {Standard}, University of Iowa, 2017.

\bibitem{presburger_completeness_1991}
Presburger M, Jacquette D.
\newblock On the completeness of a certain system of arithmetic of whole
  numbers in which addition occurs as the only operation.
\newblock \emph{History and Philosophy of Logic}, 1991.
\newblock \textbf{12}(2):225--233.
\newblock \doi{10.1080/014453409108837187}.

\bibitem{ginsburg_semigroups_1966}
Ginsburg S, Spanier E.
\newblock Semigroups, {Presburger} formulas, and languages.
\newblock \emph{Pacific journal of Mathematics}, 1966.
\newblock \textbf{16}(2):285--296.
\newblock \doi{10.2140/pjm.1966.16.285}.

\bibitem{hauschildt_semilinearity_1990}
Hauschildt D.
\newblock Semilinearity of the reachability set is decidable for {Petri} nets.
\newblock {PhD} {Thesis}, University of Hamburg, Germany, 1990.

\bibitem{lambert_vector_1990}
Lambert JL.
\newblock Vector addition systems and semi-linearity.
\newblock Universit\'e Paris-Nord. Centre Scientifique et Polytechnique [CSP],
  1990.

\bibitem{finkel_how_2002}
Finkel A, Leroux J.
\newblock How to {Compose} {Presburger}-{Accelerations}: {Applications} to
  {Broadcast} {Protocols}.
\newblock In: {Foundations} of {Software} {Technology} and {Theoretical}
  {Computer} {Science} (FSTTCS), volume 2556 of \emph{Lecture Notes in Computer
  Science}. Springer, 2002 \doi{10.1007/3-540-36206-1_14}.

\bibitem{reductron}
Amat N.
\newblock Reductron: {The} {Polyhedral} {Abstraction} {Prover}. {A} tool to
  automatically prove the correctness of polyhedral equivalences for {Petri}
  nets., 2023.
\newblock \urlprefix\url{https://github.com/nicolasAmat/Reductron}.

\bibitem{laas-cnrs_file_2020}
{LAAS-CNRS}.
\newblock File formats of the {Tina} {Toolbox}.
\newblock \urlprefix\url{http://projects.laas.fr/tina//manuals/formats.html/}.

\bibitem{de2008z3}
De~Moura L, Bj{\o}rner N.
\newblock Z3: {An} {Efficient} {SMT} {Solver}.
\newblock In: Tools and {Algorithms} for the {Construction} and {Analysis} of
  {Systems} ({TACAS}), volume 4963 of \emph{Lecture Notes in Computer Science}.
  Springer, 2008 \doi{10.1007/978-3-540-78800-3_24}.
\end{thebibliography}


 \end{document}